\documentclass[%
 aip,
 sd,%
 amsmath,amssymb,
 preprint,%
]{revtex4-1}

\usepackage{graphicx}
\usepackage{color}
\usepackage{dcolumn}
\usepackage{bm}
\usepackage{verbatim} 

\begin{document}

\long\def \cblu#1#2{{\color{blue}#1 }{\color{red}(#2)}}

    \long\def \cblu#1{\color{blue}#1}
    \long\def \cred#1{\color{red}#1}
    \long\def \cgre#1{\color{green}#1}
    \long\def \cpur#1{\color{purple}#1}

\def\FileRef{
\input FName
{
\newcount\hours
\newcount\minutes
\newcount\min
\hours=\time
\divide\hours by 60
\min=\hours
\multiply\min by 60
\minutes=\time
\
\advance\minutes by -\min
{\small\rm\em\the\month/\the\day/\the\year\ \the\hours:\the\minutes
\hskip0.125in{\tt\FName}
}
}}

\mathchardef\muchg="321D
\let\na=\nabla
\let\pa=\partial

\let\muchg=\gg

\let\t=\tilde
\let\ga=\alpha
\let\gb=\beta
\let\gc=\chi
\let\gd=\delta
\let\gD=\Delta
\let\ge=\epsilon
\let\gf=\varphi
\let\gg=\gamma
\let\gh=\eta
\let\gj=\phi
\let\gF=\Phi
\let\gk=\kappa
\let\gl=\lambda
\let\gL=\Lambda
\let\gm=\mu
\let\gn=\nu
\let\gp=\pi
\let\gq=\theta
\let\gr=\rho
\let\gs=\sigma
\let\gt=\tau
\let\gw=\omega
\let\gx=\xi
\let\gy=\psi
\let\gY=\Psi
\let\gz=\zeta

\let\lbq=\label
\let\rfq=\ref
\let\na=\nabla
\def\daI{{\dot{I}}}
\def\dsq{{\dot{q}}}
\def\dgj{{\dot{\phi}}}

\def\bgs{\bar{\sigma}}
\def\bgh{\bar{\eta}}
\def\bgg{\bar{\gamma}}
\def\bgF{\bar{\Phi}}
\def\bgY{\bar{\Psi}}

\def\baF{\bar{F}}
\def\bsj{\bar{j}}
\def\baJ{\bar{J}}
\def\bsp{\bar{p}}
\def\bsx{\bar{x}}

\def\hgj{\hat{\phi}}
\def\hgq{\hat{\theta}}

\def\HaT{\hat{T}}
\def\HaR{\hat{R}}
\def\Hsb{\hat{b}}
\def\Hsh{\hat{h}}
\def\Hsz{\hat{z}}

\let\gG=\Gamma
\def\taA{{\tilde{A}}}
\def\taB{{\tilde{B}}}
\def\taG{{\tilde{G}}}
\def\tsp{{\tilde{p}}}
\def\tsv{{\tilde{v}}}
\def\tgF{{\tilde{\Phi}}}

\def\wgx{{\bm{\xi}}}
\def\wgz{{\bm{\zeta}}}

\def\wse{{\bf e}}
\def\wsk{{\bf k}}
\def\wsi{{\bf i}}
\def\wsj{{\bf j}}
\def\wsl{{\bf l}}
\def\wsn{{\bf n}}
\def\wsr{{\bf r}}
\def\wsu{{\bf u}}
\def\wsv{{\bf v}}
\def\wsx{{\bf x}}

\def\vaB{\vec{B}}
\def\vse{\vec{e}}
\def\vsh{\vec{h}}
\def\vsl{\vec{l}}
\def\vsv{\vec{v}}
\def\vgn{\vec{\nu}}
\def\vgk{\vec{\kappa}}
\def\vgt{\vec{\gt}}
\def\vgx{\vec{\xi}}
\def\vgz{\vec{\zeta}}

\def\waA{{\bf A}}
\def\waB{{\bf B}}
\def\waE{{\bf E}}
\def\waJ{{\bf J}}
\def\waV{{\bf V}}
\def\waX{{\bf X}}

\def\R#1#2{\frac{#1}{#2}}
\def\btbl{\begin{tabular}}
\def\etbl{\end{tabular}}
\def\bqbl{\begin{eqnarray}}
\def\eqbl{\end{eqnarray}}
\def\ebox#1{
  \begin{eqnarray}
    #1
\end{eqnarray}}


\def \cred#1{{\color{red}(#1)}}
\def \cblu#1{{\color{blue}#1}}

\preprint{AIP/123-QED}

\title{Energy spectrum of tearing mode turbulence in sheared background field}
\author{Di Hu}
 \altaffiliation[While visiting at ]{PPPL, Princeton, New Jersey}
 \email{hudi\_2@pku.edu.cn}
\affiliation{
School of Physics, Peking University, Beijing 100871, China.
}
\affiliation{
ITER Organization, Route de Vinon sur Verdon, CS 90 046,13067 Saint Paul-lez-Durance, Cedex, France.
}

\author{Amitava Bhattacharjee}
\author{Yi-Min Huang}
\affiliation{
Department of Astrophysical Sciences, Princeton University, Princeton,
New Jersey, 08544, USA
}
\affiliation{
Princeton Plasma Physics Laboratory, Princeton University, Princeton,
New Jersey, 08540, USA
}

\date{\today}

\begin{abstract}
	The energy spectrum of tearing mode turbulence in a sheared background magnetic field is studied in this work. We consider the scenario where the nonlinear interaction of overlapping large-scale modes excites a broad spectrum of small-scale modes, generating tearing mode turbulence. The spectrum of such turbulence is of interest since it is relevant to the small-scale back-reaction on the large-scale field. The turbulence we discuss here differs from traditional MHD turbulence mainly in two aspects. One is the existence of many linearly stable small-scale modes which cause an effective damping during energy cascade. The other is the scale-independent anisotropy induced by the large-scale modes tilting the sheared background field, as opposed to the scale-dependent anisotropy frequently encountered in traditional critically balanced turbulence theories. Due to these two differences, the energy spectrum deviates from a simple power law and takes the form of a power law multiplied by an exponential falloff. Numerical simulations are carried out using visco-resistive MHD equations to verify our theoretical predictions, and reasonable agreement is found between the numerical results and our model.
\end{abstract}


\maketitle

\section{Introduction}
\label{s:Intro}

\vskip1em

The generation of a spectrum of small-scale tearing modes by their large-scale counterparts is a very relevant
issue both in magnetically confined devices such as a reversed-field-pinch
(RFP) or a tokamak as well as in space and astrophysical plasmas. For RFPs, the constant
interaction of tearing modes and resistive interchange modes keeps the
plasma in a perpetual turbulent state \cite{RFP}. For tokamaks, the
non-linear excitation and overlapping of a spectrum of tearing modes may break nested flux surfaces and lead to
disruption \cite{Diamond1984,Strauss1986,Craddock1991}. In astrophysical
plasmas, secondary plasmoid turbulence is found to play a crucial
role during magnetic reconnection both in kinetic \cite{Daughton2011}
and in resistive MHD\cite{Huang16APJ} investigations.

An important aspect of these problems is the back-reaction of
small-scale field fluctuations on their large-scale
counterparts. A well-known example of such back-reaction is the
hyper-resistivity produced in a mean-field theory, which has been a subject of intensive studies in the past decades
\cite{Strauss1986,Craddock1991,Boozer1986,AB1986,Hameiri1987}. To
understand this problem, however, knowledge regarding the structure of tearing turbulence spectrum is necessary
\cite{Diamond1984,Strauss1986,Craddock1991,AB1986,Hameiri1987}. Hence, in this paper, we try to construct a model to describe the structure of tearing-instability-driven turbulence spectrum in a sheared strong magnetic field. While this sheared and strongly magnetized case would appear to be most relevant to laboratory plasmas and to space and astrophysical plasmas characterized by strong guide fields, our approach also provides important qualitative insight into more general problems where turbulence is instability-driven due to strong spatial inhomogeneities.

Two arguments are commonly invoked when studying the spectrum of MHD turbulence. One is the inertial range argument,
which states that there exists a self-similar region in the $k$ space between the energy injection scale and dissipation scale where energy is
conservatively transferred from one scale to another,
resulting in a power-law energy spectrum\cite{FrischBook,DiamondBook}. The other is the scale-dependent anisotropy which indicates that the ratio between the
parallel and perpendicular length scale $l_\|/l_\bot$ of turbulent eddies depends on
$l_\bot$. For weak turbulence in a homogeneous magnetic field, three-wave
interactions result in no cascade along the parallel direction
\cite{Ng96ApJ,Ng96POP,Galtier2000,Lithwick2003}. Hence,
$l_\|$ is independent of $l_\bot$, which yields an energy spectrum
$E\left(k_\bot,l_\|\right)=E_\bot\left(k_\bot\right)f\left(l_\|\right)\propto
k_\bot^{-2}$, where $f\left(l_\|\right)$ is any initial spectrum function of $l_\|$ and
$k_\bot\sim l_\bot^{-1}$ is the perpendicular wave number.
For strong turbulence, assuming no scale-dependent alignment, the frequently
invoked critical balance condition assumes that the nonlinear term and linear
term are of the same order, $v_A/l_\|\sim v\left(l_\bot\right)/l_\bot$, where
$v_A$ is the Alfv\'{e}n speed of the background field and $v\left(l_\bot\right)$ is
the velocity at a given perpendicular scale $l_\bot$ \cite{GS1995,GS1997}. Combining the
critical balance assumption with the inertial range argument yields the
scale-dependent anisotropy $l_\|\propto l_\bot^{2/3}$, corresponding to
the energy spectrum $E\left(k_\bot\right)\propto k_\bot^{-5/3}$ \cite{DiamondBook}.
With scale-dependent alignment, the balance between linear and nonlinear terms
becomes $v_A/l_\|\sim v^2\left(l_\bot\right)/v_A l_\bot$, leading to the anisotropy relation
$l_\|\propto l_\bot^{1/2}$, and the energy spectrum $E\left(k_\bot\right)
\propto k_\bot^{-3/2}$ \cite{Boldyrev2006}.

However, recent development in kinetic turbulence theory has pointed out
the possibility that stable eigenmodes nonlinearly excited by unstable modes can act as an effective damping mechanism \cite{HatchPRL2011,HatchPOP2011}. This is equally true for tearing
turbulence with which we are concerned here. Unlike the commonly discussed externally driven
turbulence in a homogeneous system, instability
driven turbulence usually has many stable modes along with a few
unstable modes which provide the energy for the rest of the spectrum. The
effective damping caused by the stable modes interrupt the transfer of energy between scales and thus alter the structure
of the spectrum. It may then be expected that the resulting
spectrum will deviate from the traditional power-law form
$E\left(k_\bot\right)\propto k_\bot^{\gb_0}$ and take the form of a power law
multiplied by an exponential fall
$E\left(k_\bot\right)\propto k_\bot^{\gb_1} \exp{\left(-\gd k_\bot^{\gb_2}\right)}$
\cite{Terry2009,Terry2012}. Here, $\gb_0$, $\gb_1$, $\gd$ and $\gb_2$ are constant
coefficients. Furthermore, a recent resistive MHD simulation concerning
plasmoid-mediated turbulence in a sheared magnetic field has found discrepancy from
the scale-dependent anisotropy picture and produced an approximately scale-independent anisotropy $l_\|\propto l_\bot$ in strong turbulence when the magnitude of the magnetic field perturbation is comparable with that of the background field
\cite{Huang16APJ}. These results raise doubt regarding the validity of the standard inertial range picture as well as that of scale-dependent anisotropy for
tearing mode turbulence in a magnetically sheared system.

In the light of the discussion above, in this paper we revisit the problem of the spectrum of tearing mode turbulence. On one hand, the presence of large-scale perturbations
in a sheared guide field is found to introduce a scale-independent anisotropy in
the small-scale eddies. On the other hand, we find significant effective damping of the
turbulence calculated from linear stability of high $k_\bot$ modes,
wherein the effective damping scales as $k_\bot^p$,
with $p=6/5$ and $4/3$ in the inviscid and viscous regime, respectively.
This effective damping has a considerably weaker dependence on $k_\bot$ than that of
classical dissipation, which generally scales as $k_\bot^2$.
We provide an analytical model for turbulence under such
scale-independent anisotropy and effective
damping. Based on this model, the modified spectrum will be obtained by
considering the local energy budget in $k$ space. This
analytical spectrum will then be compared with resistive MHD
simulation. Reasonable agreement is found between analytical predictions and numerical results.

The rest of the paper is arranged as follows. In Section \ref{s:System},
the system of interest will be described and the basic resistive MHD
equations will be introduced. In Section \ref{s:DampedTurbulence}, the
theoretical model regarding the damped tearing turbulence and the modified turbulence spectrum will be discussed.
This new spectrum
will be checked with simulation results in Section \ref{s:Simulation},
and spectral properties as well as structure functions of the turbulence will be discussed. The turbulence anisotropy will be studied analytically as well as numerically.
Furthermore, this scale-independent anisotropy will be checked for strong turbulence
cases. Discussions on the implication of this new form of spectrum
to future studies and a conclusion will be presented in Section
\ref{s:Conclusion}.

\section{System of interest}
\label{s:System}

\vskip1em

We will consider the standard compressible MHD equations with viscosity
and resistivity included, as follows:
\bqbl
\lbq{eq:Continuous}
\R{\pa}{\pa t}\gr
+\na\cdot\left(\gr \wsv\right)
=
0
,\eqbl
\bqbl
\lbq{eq:Motion}
\R{\pa}{\pa t}\left(\gr\wsv\right)
+\gr\left(\wsv\cdot\na\right)\wsv
+\wsv\left[\na\cdot\left(\gr\wsv\right)\right]
=
-\na\left(p+\R{B^2}{2}\right)
+\left(\waB\cdot\na\right)\waB
+\gn\na^2\left(\gr\wsv\right)
,\eqbl
\bqbl
\lbq{eq:State}
\R{\pa}{\pa t}p
+\na\cdot\left(p\wsv\right)
=
-\left(\gg_A-1\right)p\na\cdot\wsv
,\eqbl
\bqbl
\lbq{eq:Ohm}
\R{\pa}{\pa t}\waB
=
\na\times\left(\wsv\times\waB-\gh\waJ\right)
.\eqbl
Here, Eq.\,(\rfq{eq:Continuous}) is the continuity equation,
Eq.\,(\rfq{eq:Motion}) is the equation of motion, Eq.\,(\rfq{eq:State})
represents the equation of state, and Eq.\,(\rfq{eq:Ohm}) is the Ohm's
law. Here $\gr$ is the plasma density, $\wsv$ is the velocity, $\waB$ is the
total magnetic field, $\waJ$ is the current density, and $p$ is the
pressure. The vacuum permeability $\gm_0$ has been absorbed into $\gr$ and $\waJ$.
Furthermore, $\gg_A=5/3$ here is the adiabatic index (which should not be
confused with the growth rate of the tearing modes). The constant dissipation
coefficients $\gn$ and $\gh$ stand for classic viscosity and resistivity
respectively.

In this study, we will consider a simple slab system with coordinates
$\left(x,y,z\right)$, and the boundary conditions are assumed to be
periodic at all sides. The sizes of the system in $x$, $y$, $z$ directions are $X$, $Y$ and $Z$ respectively, and the geometric center of the system is chosen to be
$\left(x,y,z\right)=\left(0,0,0\right)$.
The three components of the equilibrium magnetic field $\waB_0$ are the following:
\bqbl
B_{x0}
=
0
,\quad
B_{y0}
=
B_{y0}\left(0\right)\cos{\left(\R{2\gp}{X}x\right)}
,\quad
B_{z0}
=
\sqrt{B_{0}^2-B_{y0}^2}
.\eqbl
Here, $B_{y0}\left(0\right)$ and $B_0$ are constants to be specified later.
The system is initially in force-free, with the pressure assumed to be constant
and set to unity.
The corresponding initial current profile is
\bqbl
J_{z0}
=
-B_{y0}\left(0\right)\R{2\gp}{X}\sin{\left(\R{2\gp}{X}x\right)}
.\eqbl
An artificial constant electric field along $z$ direction is implemented
to sustain the initial current profile against resistive diffusion.
Although the assumed global geometry is simple, it is sufficient to capture the fundamental physical process of the dynamics of small-scale tearing fluctuations. The qualitative features of the theory are not expected to change in more realistic global geometry.

\begin{figure*}
\centering
\noindent
\btbl{c}
\parbox{5.0in}{
  \includegraphics[scale=0.5]{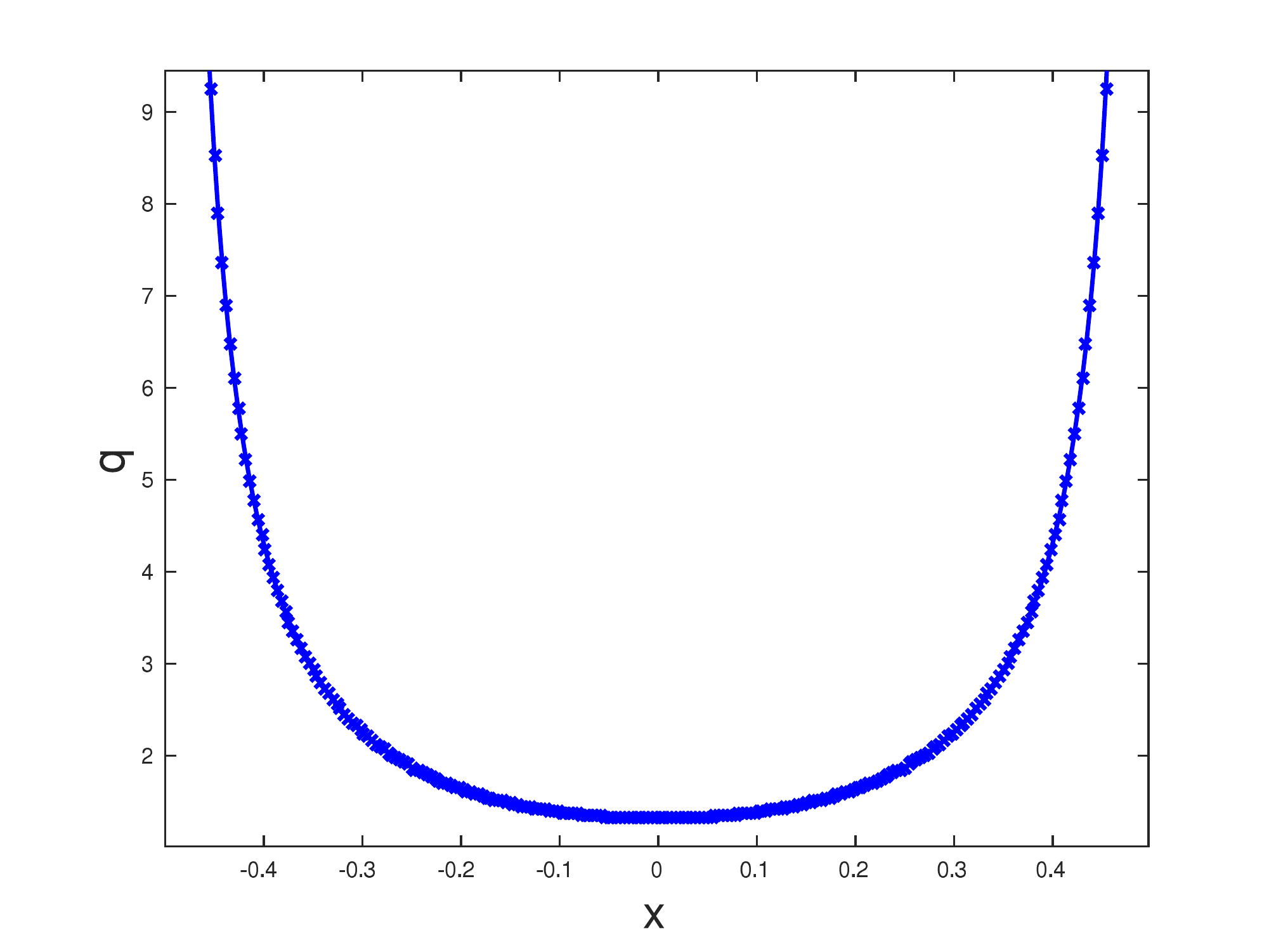}
}
\etbl
\caption{The analogous safety factor profile in region
  $x\in\left[-0.5,0.5\right]$ for initial background magnetic field,
  with $X=2$, $Y=4$, $Z=20$, $B_{0}=10$ and
  $B_{y0}\left(0\right)=1.5$. The safety factor tends to infinity near
  $x=\pm0.5$ due to $B_{y0}$ being zero there.
}
\label{fig:1}
\end{figure*}

We define a ``safety factor''
\bqbl
q
\equiv
\R{Y B_{z0}}{Z B_{y0}}
\eqbl
and the ``rotational transform''
\bqbl
\gm
\equiv
\R{1}{q}
\eqbl
analogous to that of a tokamak. The corresponding $q$ profile is then a function of $x$. In the region
$x\in\left[-0.5,0.5\right]$, the $q$ profile is shown in
Fig.\,\ref{fig:1}, with $X=2$, $Y=4$, $Z=20$, $B_{0}=10$,
$B_{y0}\left(0\right)=1.5$, and the corresponding minimum safety factor is given by
$q\left(0\right)=1.3$. Numerical observation indicates that several
large-scale modes, such as $2/1$, $3/1$ and $3/2$ modes, are unstable
for this magnetic shear profile. The nonlinear growth and interaction of these
modes will then generate a spectrum of small-scale modes.

As the turbulence grows in strength, it will have a back-reaction on
the mean background field, leading to self-consistent evolution of the
latter. The mean
current profile will tend to relax under turbulence spreading
\cite{Diamond1984,Strauss1986,Craddock1991}, and it is
observed that substantial profile flattening would occur over time after
the turbulence has been fully established. Ultimately,
the relaxation would reach a point where there is no free energy
available, and the tearing turbulence would then gradually decay away.
However, it will be shown in Section \ref{s:StrongField} that the
characteristic time scale of such decay is much longer than the slowest
nonlinear turnover time of eddies, thus the turbulence can be viewed as having
attained a quasi-steady-state before decay occurs.

\section{Analytical model for tearing turbulence}
\label{s:DampedTurbulence}

\vskip1em

The structure of tearing turbulence spectrum will be discussed analytically in this
section. Three quantities are needed in order to obtain the spectrum of tearing turbulence in a sheared guide field. The first is the effective damping rate caused by small-scale linearly stable modes, the second is the anisotropy property of the tearing turbulence, and the third is the local energy transfer in the $k$ space \cite{FrischBook,DiamondBook}. We will treat the effective damping and anisotropy property in Section \ref{s:EffectiveDamping} and \ref{s:Anisotropy} respectively, then substitute these results into the local energy transfer equation in Section \ref{s:EnergyBudget} to obtain the turbulence spectrum. In Section \ref{s:EffectiveDamping}, we will first justify the use of linear stability theory in considering the effective damping, then provide the $k_\bot$ scaling of growth rate and further obtain the effective damping rate for inviscid and viscous limit in Eq.\,(\rfq{eq:EffectiveDamp})-(\rfq{eq:Coefficient2}). In Section \ref{s:Anisotropy}, we will investigate the scale-dependence of turbulence anisotropy by considering the ratio between the parallel wave number dispersion $\gD k_\|\sim l_\|^{-1}$ as defined in Eq.\,(\rfq{eq:k_para_dispersion}) and the perpendicular wave number $k_\bot\sim l_\bot^{-1}$. The result is given in Eq.\,(\rfq{eq:weakanisotropy}) and Eq.\,(\rfq{eq:scaleindep}) for unperturbed and perturbed sheared guide field respectively. Finally, in Section \ref{s:EnergyBudget}, we combine the aforementioned results with the local energy budget in Eq.\,(\rfq{eq:EBudget}) and the forward energy transfer rate in Eq.\,(\rfq{eq:Asymptote}) to obtain the spectrum shape shown in Eq.\,(\rfq{eq:Spectrum}).

\subsection{Effective damping caused by linearly stable modes}
\label{s:EffectiveDamping}

\vskip1em

We consider the effective damping under the
assumption of weak nonlinearity, that is, the nonlinear interaction is assumed to be sufficiently weak that it does not change the linear
outer region solution. Hence, we can still use linear theory to
consider the mode structure, and the effective damping rate can be
estimated from the negative linear growth rate.

The justification of using the linear growth rate to estimate effective
damping may be formulated more precisely as follows. The effective island width $w$
for a given Fourier component of the magnetic perturbation
$\taB_k\left(x,y,z\right)=\taB_k^{(0)}
\left(x\right) \exp{\left(ik_yy-ik_zz\right)}$ has the following
dependence on mode numbers and the magnetic perturbation strength:
\cite{WhiteRMP,Di2015}
\bqbl
w
\sim
\left(-\gy/\gY_{0s}''\right)^{1/2}
\sim
\left(\R{\taB_x}{B_{z0}}\R{L_s}{k_y}\right)^{1/2}
,\eqbl
where $\gy$ is the perturbed oblique flux
\bqbl
\gy\equiv\taA\cdot\vsh
,\quad
\vsh
\equiv
\vse_z
+\left(k_z/k_y\right)\vse_y
.\eqbl
Here, $\vsh$ is the oblique direction defined by given $k_y$
and $k_z$. Also, $\taB_x$ is the $x$ component of the
corresponding magnetic perturbation and $\gY_{0s}''$ is the second
order derivative of background oblique flux taken at the resonant
surface. Furthermore, $L_s\equiv
Zq/s$ is the magnetic shear length and $s\equiv Yq'/q$ is the magnetic
shear.  In the inviscid limit, the tearing layer width scales as
\cite{Rutherford1973,Coppi1966,Glasser1975}
\bqbl
x_\gh
\sim
\left(\R{\gh}{v_A}\R{L_s}{k_y}\right)^{2/5}\left(\gD'\right)^{1/5}
,\eqbl
where $\gD'\equiv \gy_s'/\gy_s\big|_{-}^{+}$ is the tearing stability index;
the minus and plus signs here denote the left and the right side of the
resonant surface. Alternatively, in the viscous regime we have \cite{Finn2005}
\bqbl
x_\gh
\sim
\left(\R{\gh}{v_A}\R{L_s}{k_y}\right)^{1/3}P_m^{1/6}
,\eqbl
where the magnetic Prandtl number $P_m\equiv \gn/\gh$.
The following two factors justify the use of linear stability analysis.
First, the perturbation
amplitudes of high-$k$ modes are orders of magnitude smaller than that
of low $k$ modes, thus the effective width of a high-$k$ island will
also be much smaller than that of a low-$k$ island. Second, the
effective island width will shrink faster than the tearing layer width
for increasing $k$, as the power dependence on $k$ for the former is
greater than that of the latter. Simple estimation using the turbulence
spectrum obtained later in Section \ref{s:Simulation} indicates that,
in our case of weak turbulence, the
island width will be smaller than the tearing layer width when
$k_\bot\geq 25$.\
Furthermore, the contribution from hyper-resistivity is also ignored since
it is proportional to the driven mode width to the fourth power, making its contribution
less important for very small-scale modes. \cite{Craddock1991}

We now examine the linear growth rate of the small-scale modes.
The ideal linear eigen-equation for slab geometry can be written as
\cite{Furth1963,BiskampBook,Baalrud2012}:
\bqbl
\lbq{eq:IdealEquation}
\pa_x^2 \gy
=
\left(k^2+\R{F''}{F}\right) \gy
.\eqbl
Here, $F\equiv \waB_0\cdot\wsk$, and $k$ is the wave number
perpendicular to the oblique direction $\vsh$. It should be noted that we
have $k_\bot\simeq k$ due to $k_\|\ll k_\bot$ as a result of the localized
small-scale mode structure. For straight tearing modes
with $k_z=0$, $F''/F$
remains finite even at the resonant surface where $F=0$. If
$k_\bot^2\muchg F''/F$, then the eigen-structure has the following form
near resonant surface $x=x_s$:
\bqbl
\lbq{eq:ModeStructure}
\gy
\simeq
\gy_s\exp{\left(-k_\bot\left|x-x_s\right|\right)}
.\eqbl
Hence, for high $k$ modes which are linearly stable, we have:
\bqbl
\lbq{eq:DeltaPrime}
\gD'
\equiv
\R{\gy_s'}{\gy_s}\Big|_{-}^{+}
\simeq
-2k_\bot
.\eqbl
For oblique modes, there is a logarithmic singularity in
the derivative of the ideal solution since $F''/F$ is singular near the
resonant surface \cite{Furth1963,Di2015}. However, the contribution
of this logarithmic singularity to
$\gy'$ is even in parity near the resonant surface, thus does not contribute to
$\gD'$. Hence, the $\gD'$ of high-$k$ oblique modes should have the same
form as that of straight modes as shown in
Eq.\,(\rfq{eq:DeltaPrime}). Numerical solution of
Eq.\,(\rfq{eq:IdealEquation}) confirms this statement
\cite{Baalrud2012}.

The linear growth rate for oblique tearing modes in the inviscid limit is given by
\cite{BiskampBook,Baalrud2012}
\bqbl
\lbq{eq:LinearGrowth1}
\gg
=
\gh^{3/5}\left(\gD'\right)^{4/5}\left(k_\bot B_{ys}'\right)^{2/5}\gr^{-1/5}
,\eqbl
while in the viscous regime we have \cite{Finn2005}
\bqbl
\lbq{eq:LinearGrowth2}
\gg
=
\gh^{2/3}P_m^{-1/6}\gD'\left(k_\bot B_{ys}'\right)^{1/3}\gr^{-1/6}
.\eqbl
Here, $B_{ys}'$ is the $x$ gradient of $B_y$ taken at resonance $x_s$.
The stable eigenmodes satisfying these dispersion relations are similar in mode structure and parity to the unstable modes that drive the turbulence. Equations (\ref{eq:LinearGrowth1}) and (\ref{eq:LinearGrowth2}) give
 the following $k_\bot$ dependence for $\gg$:
\bqbl
\gg
\propto
-\gh^{3/5}k_\bot^{6/5}
\eqbl
in the inviscid limit and
\bqbl
\gg
\propto
-\gh^{2/3}P_m^{-1/6}k_\bot^{4/3}
\eqbl
in the viscous regime.

As has been mentioned in Section \ref{s:System}, the background
magnetic field is constantly evolving throughout the time-evolution of turbulence,
hence we need to track the evolution of $B_{ys}'$ numerically as the turbulence evolves.
We define the following characteristic length scale of $B_y$
variation:
\bqbl
\gl
\equiv
\R{B_{y0}\left(0\right)}{B_{ys}'.}
\eqbl
Thus, the effective damping in $k$ space can be written as:
\bqbl
\lbq{eq:EffectiveDamp}
\left[\pa_t E\left(k\right)\right]_{damping}
=
2\gg E\left(k\right)
=
-2DS^{-p/2}\left(k_\bot\gl\right)^{p}E\left(k\right)
,\eqbl
with $p=6/5$ in the inviscid limit and $p=4/3$ in the viscous limit.
Here, $E\left(k_\bot\right)=v\left(k_\bot\right)^2/k_\bot$ is the ``energy
density'' in $k_\bot$ space. We consider \emph{a priori} the
equipartition of magnetic and kinetic energy for medium to high $k_\bot$. (We
will check the validity of this assumption \emph{a posteriori}).
The Lundquist number $S$ is defined as $S\equiv \gt^*_\gh/\gt_A$, with
$\gt^*_\gh\equiv \gl^2/\gh$ and
$\gt_A\equiv Z/v_A$, while $v_A$ is the Alfv\'{e}n speed corresponding to the guide field.
Furthermore, $D$ is the effective damping coefficient with dimension of
$1/t$. Combining Eq.\,(\rfq{eq:LinearGrowth1}) or Eq.\,(\rfq{eq:LinearGrowth2}) with
Eq.\,(\rfq{eq:EffectiveDamp}), we obtain
\bqbl
\lbq{eq:Coefficient1}
D
=
1.41\left(\R{Z}{\gl}\right)^{2/5}
\left(\R{B_{y0}}{B_{z0}}\right)^{2/5}\gt_A^{-1}
\eqbl
in the inviscid limit and
\bqbl
\lbq{eq:Coefficient2}
D
=
2\left(\R{Z}{\gl}\right)^{1/3}
\left(\R{B_{y0}}{B_{z0}}\right)^{1/3}\gt_A^{-1}P_m^{-1/6}
\eqbl
in the viscous regime.
The damping rate given in Eq.\.(\rfq{eq:EffectiveDamp}) has a weaker dependence on $k_\bot$ than the
classical dissipation does, making the distinction between the inertial
range and the dissipation range hard to define.
Thus, the present physical situation, in which damping appears to be
important at all scales, does not permit a strict delineation of an
inertial range in tearing turbulence.

\subsection{Scale-independent anisotropy in sheared background field}
\label{s:Anisotropy}

\vskip1em

The scale dependence of the ratio between the parallel and the
perpendicular length scales of eddies is of great interest since it
directly affects the nonlinear turnover rate and thus further
influences the forward energy cascade rate of turbulence. The nonlinear
turnover rate for MHD turbulence can be modeled as \cite{DiamondBook}:
\bqbl
\R{1}{\gt_{nl}}
\simeq
\R{v\left(k_\bot\right)^2}{l_\bot^2}\R{l_\|}{v_A}
.\eqbl
Here, $v\left(k_\bot\right)$ represents kinetic perturbation at $k_\bot$
scale.

For weak turbulence generated by oppositely propagating
Alfv\'{e}n waves with straight background field lines, the three-wave
interaction preserves the $k_\|$ space structure of the beating
waves, thus preventing any energy cascade along the direction parallel
to the background magnetic field \cite{Ng96ApJ,Ng96POP}.
A simple way to see this is by considering the
resonant condition of wave number and frequency for three-wave
interaction \cite{Shebalin1983,Sridhar2010}. We have:
\bqbl
\wsk_1
+\wsk_2
=
\wsk_3
,\quad
\gw^\pm_1
+
\gw^\mp_2
=
\gw^\pm_3
.\eqbl
Here, $\gw^+=v_Ak_\|$ and $\gw^-=-v_Ak_\|$ represent the angular frequencies of the forward and the backward propagating Alfv\'{e}n waves, respectively. The oppositely
propagating waves indicate that either $k_{1\|}$ or $k_{2\|}$ must be
zero to satisfy both resonance conditions for the wave number and the frequency. Hence, there is no cascade of energy along $k_\|$ and the nonlinear turnover rate scales as $\gt_{nl}\propto
v\left(k_\bot\right)^2l_\bot^{-2}$ as a result.

On the other hand,
for a spectrum of modes in a sheared guide field,
the parallel length scale $l_\|\simeq 1/\gD k_\|$,
where $\gD k_\|$ is the dispersion in parallel wave number,
and perpendicular length scale $l_\bot\simeq 1/k_\bot$, where $k_\bot$ is
the perpendicular wave number. The dispersion in parallel wave number, $\gD k_\|$, is defined as
\bqbl
\lbq{eq:k_para_dispersion}
\left(\gD k_\|\right)^2
\equiv
\left<k_\|^2\right>_{k_\bot,x}
-\left<k_\|\right>^2_{k_\bot,x}
.\eqbl
Here, $\left<f\right>_{k_\bot,x}$ represents averaging quantity $f$ over $k_\|$ for a given
$k_\bot$ and across the $\left(y,z\right)$ plane for a given $x$.
Averaging over the $\left(y,z\right)$ plane is necessary because the small-scale mode structures are very localized and we are looking at the spectrum at a specific $x$.
Within the framework of weak turbulence theory in a strong guide field where the
average field is assumed to be unperturbed, we will find a similar independence between
$l_\|$ and $l_\bot$ in the turbulence spectrum, i.e., $\gD k_\|\propto k_\bot^0$,
although the physical mechanism is somewhat different from that described above. However, it can be seen that
the inclusion of a finite large-scale perturbation will introduce an additional relationship
between $\gD k_\|$ and $k_\bot$ in the spectrum, so long as we have
$\taB_L L_s k_\bot/B_{z0} \muchg 1$,
where $\taB_L$ is the random large-scale perturbation, $L_s$ is the shear length of background guide field, and $B_{z0}$ is
the guide field along the ignorable direction. It is important to note that $k_\bot$ in this criterion
is the perpendicular wave number of the small-scale modes rather than the large-scale
perturbation. Thus, the left-hand-side of the aforementioned criterion should not be confused with the Kubo
number of the large-scale perturbation, defined as the ratio between the nonlinear and linear terms $\gk\equiv\left(\taB/B_0\right)/\left(l_\|/l_\bot\right)$.

We assume the perturbation has the
following form: $\taB_k\left(x,y,z\right)=\taB_k^{(0)}
\left(x\right) \exp{\left(ik_yy-ik_zz\right)}$, where $m\equiv
Yk_y/2\gp$ and $n\equiv Zk_z/2\gp$. For the unperturbed background field, we have:
\bqbl
k_\|
=
2\gp\left(
\R{B_{y0}}{B_0}\R{m}{Y}
-\R{B_{z0}}{B_0}\R{n}{Z}
\right)
=
k_y\R{Y}{Z}\R{B_{z0}}{B_0}\left(\gm-\R{n}{m}\right)
=
-\R{k_y\gD x}{L_s}\R{B_{z0}}{B_0}
,\quad
\gm
\equiv
1/q
.\eqbl
Again, $L_s\equiv Zq/s$, $q\equiv YB_{z0}/ZB_{y0}$, and $s\equiv Yq'/q$.
We repeat for emphasis that $B_{y0}$ and $B_{z0}$ here do not contain the
contribution of large-scale perturbation. The length $\gD x\equiv x-x_s$ represents
the distance to the resonant surface for a given $m/n$.

It will be shown later on in Section \ref{s:StrongField} that the characteristic length
scale of turbulence strength envelope is much larger than $1/k_\bot$ in cases we
are interested in, thus $\gy_s$ can be assumed to be independent of $\gD x$ for
a given $x$. Then Eq.\,(\rfq{eq:ModeStructure}) yields
\bqbl
\lbq{eq:average}
\left<f\right>_{k_\bot,x}
=
\left<
\R{\int_{-\infty}^{\infty}{\exp{\left(-2k_\bot\left|\gD x\right|\right)}fd\gD x}}
{\int_{-\infty}^{\infty}{\exp{\left(-2k_\bot\left|\gD x\right|\right)}d\gD x}}
\right>_x
.\eqbl
Note that here we integrate over $\gD x$ instead of $k_\|$ because
$dk_\|\propto d\gD x$ so long as $k_y\propto k_\bot$.
For the denominator, we have:
\bqbl
\int_{-\infty}^{\infty}{\exp{\left(-2k_\bot\left|\gD x\right|\right)}d\gD x}
=
\R{e^{2k_\bot\gD x}}{2k_\bot}\Big|_{-\infty}^{0}
-\R{e^{-2k_\bot\gD x}}{2k_\bot}\Big|_{0}^{\infty}
=
\R{1}{k_\bot}
.\eqbl
Thus we obtain:
\bqbl
\lbq{eq:kparave}
\left<k_\|\right>_{k_\bot}
=
0
,\quad
\left(\gD k_\|\right)^2
=
k_\bot\int_{-\infty}^{\infty}{\exp{\left(-2k_\bot\left|\gD x\right|\right)}
\left(\R{k_y\gD x}{L_s}\R{B_{z0}}{B_0}\right)^2d\gD x}
.\eqbl
Because the localized mode structure also implies that all the small-scale
modes which can be ``seen'' from $x$ have similar $\gm$, we can approximately write:
\bqbl
k_\bot
=
\R{B_{z0}}{B_0}k_y
+\R{B_{y0}}{B_0}k_z
,\\
\lbq{eq:inverse}
\R{Zk_z}{Yk_y}
\simeq
\gm\left(x\right)
.\eqbl
Therefore, we obtain
\bqbl
k_y
\simeq
\R{Z^2}{Z^2+Y^2\gm\left(x\right)^2}\R{B_0}{B_{z0}}k_\bot
.\eqbl
Substituting the above relationship into Eq.\,(\rfq{eq:kparave}), the parallel length scale $l_\|$
for small scale perturbations is found to be independent of $k_\bot$
\bqbl
\lbq{eq:weakanisotropy}
l_\|^{-2}
=
\left(\gD k_\|\right)^2
\simeq
\R{Z^2}{Z^2+Y^2\gm\left(x\right)^2}\R{1}{2L_s^2}
\propto
k_\bot^{0}
.\eqbl
This is similar to the weak turbulence limit discussed in Ref.\,[\onlinecite{Ng96ApJ}]
and Ref.\,[\onlinecite{Ng96POP}], although the underlying physics is quite different.

Now, let us consider the effect of a large-scale perturbation on the small-scale anisotropy. We consider the summation of several large-scale modes as a random magnetic perturbation strong enough to twist the field ``seen'' by the small-scale modes. Let $\taB_L$ be the perturbation component in
$\left(y,z\right)$ plane.
Thus, the parallel wave number for each mode is now:
\bqbl
\lbq{eq:paraknew}
k_\|
=
-\R{k_y\gD x}{L_s}\R{B_{z0}}{B_0}
+\R{\taB_{Ly}}{B_0}k_y
-\R{\taB_{Lz}}{B_0}k_z
.\eqbl
Recalling Eq.\,(\rfq{eq:inverse}), for given $x$, we have:
\bqbl
k_\|
\simeq
-\R{B_{z0}}{B_0}\R{k_y\gD x}{L_s}
+\R{\taB_{Ly}}{B_0}\left[1-\R{Y}{Z}\gm\left(x\right)\R{\taB_{Lz}}{\taB_{Ly}}\right]k_y
.\eqbl
For simplicity, we define the following parameters:
\bqbl
T
\equiv
\R{B_{z0}}{B_0}
,\quad
U
\equiv
\R{\taB_{Ly}}{B_0}\left[1-\R{Y}{Z}\gm\left(x\right)\R{\taB_{Lz}}{\taB_{Ly}}\right]
.\eqbl
An important feature of the latter parameter is that the contribution from the large-scale perturbation
vanishes upon taking the $\left(y,z\right)$ plane average since $U$ vanishes under such spatial
average, although $U^2$ does not.

Carrying out the same method used above, we also obtain
\bqbl
\left<k_\|\right>_{k_\bot,x}
=
\left<U\R{B_0}{B_{z0}}k_\bot\right>_x
=
0
,\eqbl
as well as
\bqbl
\left<k_\|^2\right>_{k_\bot,x}
&
\simeq
&
\left<
k_\bot\R{B_0^2}{B_{z0}^2}\int_{-\infty}^{\infty}
{
e^{-2k_\bot\gD x}\left[
T^2\left(\R{k_\bot}{L_s}\right)^2\gD x^2
-2TU\R{k_\bot^2}{L_s}\gD x
+U^2k_\bot^2
\right]d\gD x
}
\right>_x
\nonumber
\\
&
=
&
\left<
\R{B_0^2}{B_{z0}^2}
\left(
\R{T^2}{2L_s^2}
+U^2k_\bot^2
\right)
\right>_x
.\eqbl
Thus, so long as $2U^2k_\bot^2L_s^2/T^2\muchg 1$, we have
\bqbl
\lbq{eq:scaleindep}
\gD k_\|
=
\sqrt{\left<
\R{B_0^2}{B_{z0}^2}
\left(
\R{T^2}{2L_s^2}
+U^2k_\bot^2
\right)
\right>_x}
\simeq
\R{B_0}{B_{z0}}
\sqrt{\left<U^2\right>_x}k_\bot
\propto
k_\bot
,\eqbl
resulting in a scale-independent anisotropy $l_\|/l_\bot\propto l_\bot^0$.
Here, we emphasize that $U/T$ can still be small for the condition $2U^2k_\bot^2L_s^2/T^2\muchg 1$ to be valid due to the largeness of $L_sk_\bot$, with $k_\bot$ being the wave number of small-scale modes.

\subsection{Local energy budget in $k_\bot$ space}
\label{s:EnergyBudget}

\vskip1em

The impact of non-negligible dissipation on the structure of the spectrum
has been studied by considering the local energy budget in the $k$ space
\cite{Terry2009,Terry2012}. We will follow this methodology here, albeit in the context of turbulence with scale-independent
anisotropy instead of turbulence that is critically balanced one, as discussed in
Section \ref{s:Anisotropy}.

Under the local interaction assumption, the local energy budget in the $k_\bot$-space naturally arises from considerations of the effective damping and
classical resistive diffusion, \cite{Terry2009,Terry2012}:
\bqbl
\lbq{eq:EBudget}
-2D S^{-p/2} \left(k_\bot\gl\right)^p E\left(k_\bot\right)
-2\gt_\gh^{*-1}\left(k_\bot\gl\right)^2E\left(k_\bot\right)
=
\R{d T\left(k_\bot\right)}{dk_\bot}
,\eqbl
where $\gt_\gh^*\equiv \gl^2/\gh$, $T\left(k_\bot\right)$ is the energy
forward transfer rate at scale $k_\bot$, and
$E\left(k_\bot\right)=v\left(k_\bot\right)^2/k_\bot$ is the energy
density in the $k_\bot$ space. The energy budget
Eq.\,(\rfq{eq:EBudget}) can be solved to yield the energy spectrum
if $T\left(k_\bot\right)$ can be written as a
function of $E\left(k_\bot\right)$. The traditional scaling for MHD turbulence
without scale-dependent alignment indicates that \cite{DiamondBook}
\bqbl
\lbq{eq:Transfer}
T\left(k_\bot\right)
=
\R{v\left(k_\bot\right)^2}{\gt_{nl}}
\simeq
\R{v\left(k_\bot\right)^4}{l_\bot^2}\R{l_\|}{v_A}
.\eqbl
Here, $v\left(k_\bot\right)$ represents the kinetic perturbation at
$k_\bot$ scale, and $v_A$ is the Alfv\'{e}n speed measured with the
guide field. Due to the equipartition of kinetic and magnetic energy,
$T\left(k_\bot\right)$ also represents the forward cascade of magnetic
energy as
$v\left(k_\bot\right)=v_A\left(\taB\left(k_\bot\right)/B_0\right)$ with
$\taB\left(k_\bot\right)$ as the magnetic perturbation at $k_\bot$ scale.

Using the scale-independent anisotropy discussed before,
the forward transfer rate is now
\bqbl
\lbq{eq:Forward}
T\left(k_\bot\right)
=
\R{v\left(k_\bot\right)^4}{l_\bot\ga v_A}
,\quad
\ga
\equiv
\R{l_\bot}{l_\|}
.\eqbl
Here, $\ga$ is a constant characterizing the scale-independent
anisotropy, the value of which will be extracted from numerical
simulations of tearing turbulence.
We follow the closure technique used in Refs.\,[\onlinecite{Terry2009}] -
\,[\onlinecite{TLBook}], and write the
forward energy transfer rate
\bqbl
\lbq{eq:Asymptote}
T\left(k_\bot\right)
\,&
=
\,&
v\left(k_\bot\right)^4k_\bot\left(\ga v_A\right)^{-1}
\nonumber
\\
&
=
&
E\left(k_\bot\right)k_\bot^2\left(\ga v_A\right)^{-1}v\left(k_\bot\right)^2
\nonumber
\\
&
=
&
E\left(k_\bot\right)\ge^{1/2}k_\bot^{3/2}\left(\ga v_A\right)^{-1/2}
.\eqbl
Here, we have used the closure $v\left(k_\bot\right)^2\simeq
\ge^{1/2}k_\bot^{-1/2}\left(\ga v_A\right)^{1/2}$. This closure technique effectively builds the
inertial power-law behavior into the energy spectrum as an asymptote in the
low-damping limit. Therefore, as can be seen later in this section, the
spectrum approaches a simple power law when the effective damping vanishes.

Substituting Eq.\,(\rfq{eq:Asymptote}) into Eq.\,(\rfq{eq:EBudget}), we obtain
a linear first order ordinary differential equation for the energy spectrum:
\bqbl
\R{d}{dk_\bot}E\left(k_\bot\right)k_\bot
=
-\R{3}{2}E\left(k_\bot\right)
-\left[
2D S^{-p/2} \left(k_\bot\gl\right)^{p-1/2}
+2\gt_\gh^{*-1}\left(k_\bot\gl\right)^{3/2}
\right]
\nonumber
\\
\times
E\left(k_\bot\right)\ge^{-1/2}\left(\ga v_A\right)^{1/2}\gl^{1/2}
.\eqbl
The tearing turbulence spectrum with linear stabilities act as
effective damping is then:
\bqbl
\lbq{eq:Spectrum}
E\left(k_\bot\right)
\,&
\sim
\,&
\left(\ga v_A\right)^{1/2}\ge^{1/2}k_\bot^{-3/2}\exp{\left[
-2D S^{-p/2}\R{\ge^{-1/2}\gl^{1/2}\left(\ga v_A\right)^{1/2}}{p-\R{1}{2}}
\left(k_\bot\gl\right)^{p-1/2}
\right]}
\nonumber
\\
&
&
\times
\exp{\left[
-2\gt_\gh^*\R{\ge^{-1/2}\gl^{1/2}\left(\ga v_A\right)^{1/2}}{3/2}
\left(k_\bot\gl\right)^{3/2}
\right]}
,\eqbl
with $p=6/5$ and $p=4/3$ in the inviscid and viscous regime respectively,
while effective damping coefficient $D$ is given by Eq.\,(\rfq{eq:Coefficient1}) and Eq.\,(\rfq{eq:Coefficient2}).
From Eq.\,(\rfq{eq:Spectrum}), it can be seen that the primary impact of
effective damping is an exponential multiplier on the original power
law. In the limit of small effective damping, the spectrum recovers the
simple power-law behavior predicted by the assumption of an inertial range. Also, the
power-law behavior in the no damping limit tends to be $k_\bot^{-3/2}$
due to the scale independent anisotropy $l_\|\propto l_\bot$.

\section{Simulations}
\label{s:Simulation}

\vskip1em

In this section, the analytical result from Section
\ref{s:DampedTurbulence} will be tested against resistive MHD
simulation using the same set of equations
(\rfq{eq:Continuous})\,-\,(\rfq{eq:Ohm})
described in Section \ref{s:System}.
Specifically,we are concerned with the structure of the energy spectrum
$E\left(k_\bot\right)$ and the dependence of $k_\bot/k_\|$ on
$k_\bot$.
We will first examine the strong guide field case where
$B_{y0}/B_{z0}\sim\mathcal{O}\left(10^{-1}\right)$ and
$\taB_L/B_{z0}\sim\mathcal{O}\left(10^{-2}\right)$. We will then consider the case of
comparable guide field where $B_{y0}/B_{z0}\sim\mathcal{O}\left(1\right)$
and the large-scale perturbed field is only one order of magnitude smaller
than the guide field $\taB_L/B_{z0}\sim\mathcal{O}
\left(10^{-1}\right)$.
In the latter case, the Kubo number for the large-scale perturbation is $\gk=\left(\taB_L/B_{0}\right)
\left(L_\|/L_\bot\right)\geq 1$, corresponding to the regime where
turbulent shearing is comparable with parallel propagation.
Here, the Kubo number $\gk$ is equivalent to the $\gc$ used by Goldreich
and Sridhar in Ref.\,[\onlinecite{GS1997}].
Numerical observation of the magnetic shear length indicates we have
$2\taB_L^2L_s^2k_\bot^2/B_{z0}^2 > \mathcal{O}\left(10^2\right)$
when $k_\bot>20$ for both of the above cases.
The numerical algorithm will follow that of
Ref.\,[\onlinecite{Huang16APJ}] and Ref.\,[\onlinecite{Guzdar1993}].
Five-point finite difference
scheme is used to calculate derivatives, and trapezoidal leapfrog is
used for time stepping scheme. The resistivity is set to be
$\gh=1\times 10^{-4}$, and the magnetic Prandtl number is
$P_m\equiv\gn/\gh=1$, and thus we are in the viscous regime discussed above. In the numerical scheme, an additional artificial fourth-order dissipation
is also implemented to damp small-scale fluctuations at grid size.
This should not be confused with the real
hyper-dissipation self-consistently generated by the nonlinear
terms \cite{Strauss1986,Craddock1991}.

\subsection{Strong guide field case}
\label{s:StrongField}

\vskip1em

\begin{figure*}
\centering
\noindent
\btbl{cc}
\parbox{3.1in}{
  \includegraphics[scale=0.430]{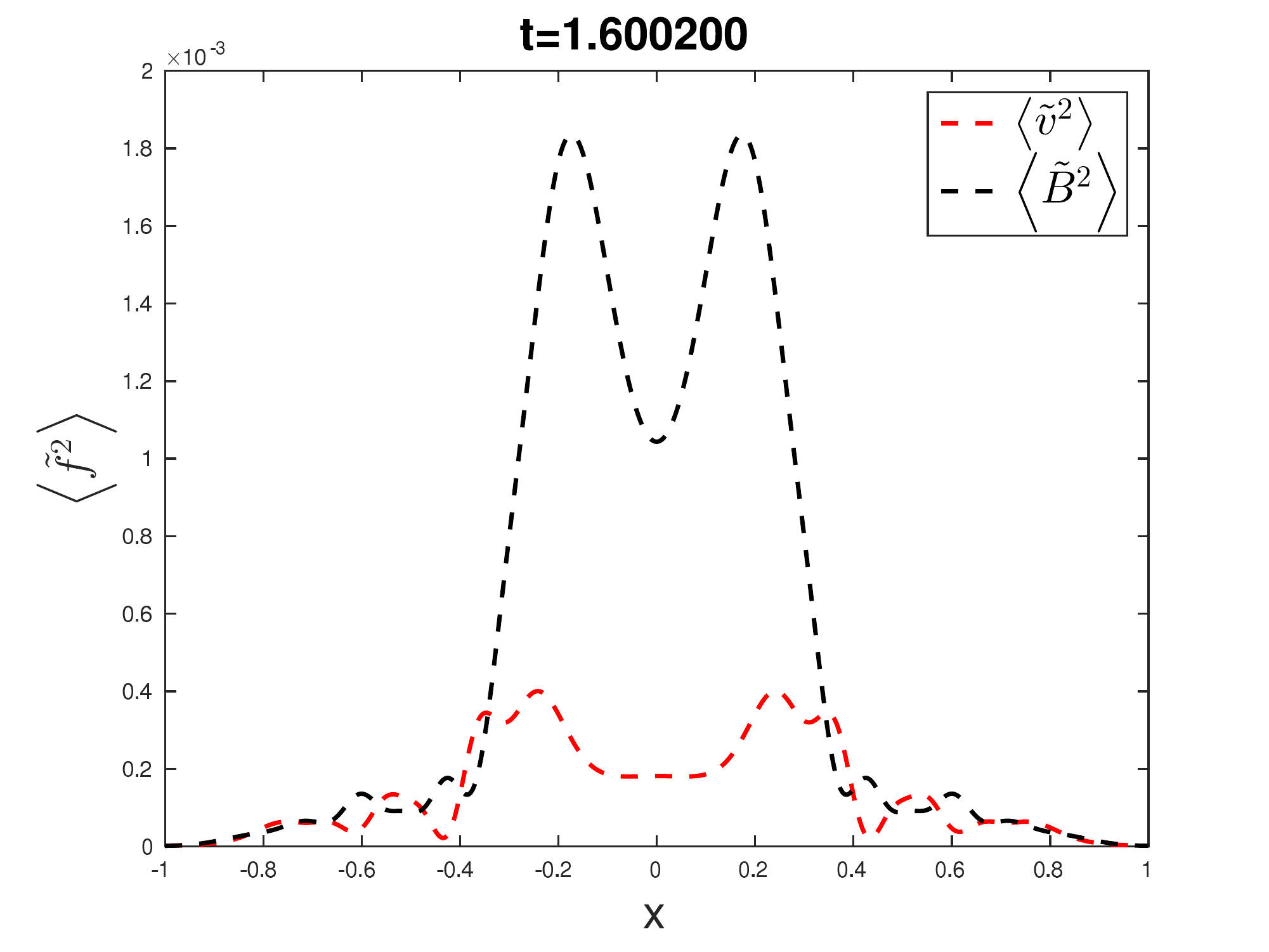}
}
&
\parbox{3.1in}{
  \includegraphics[scale=0.430]{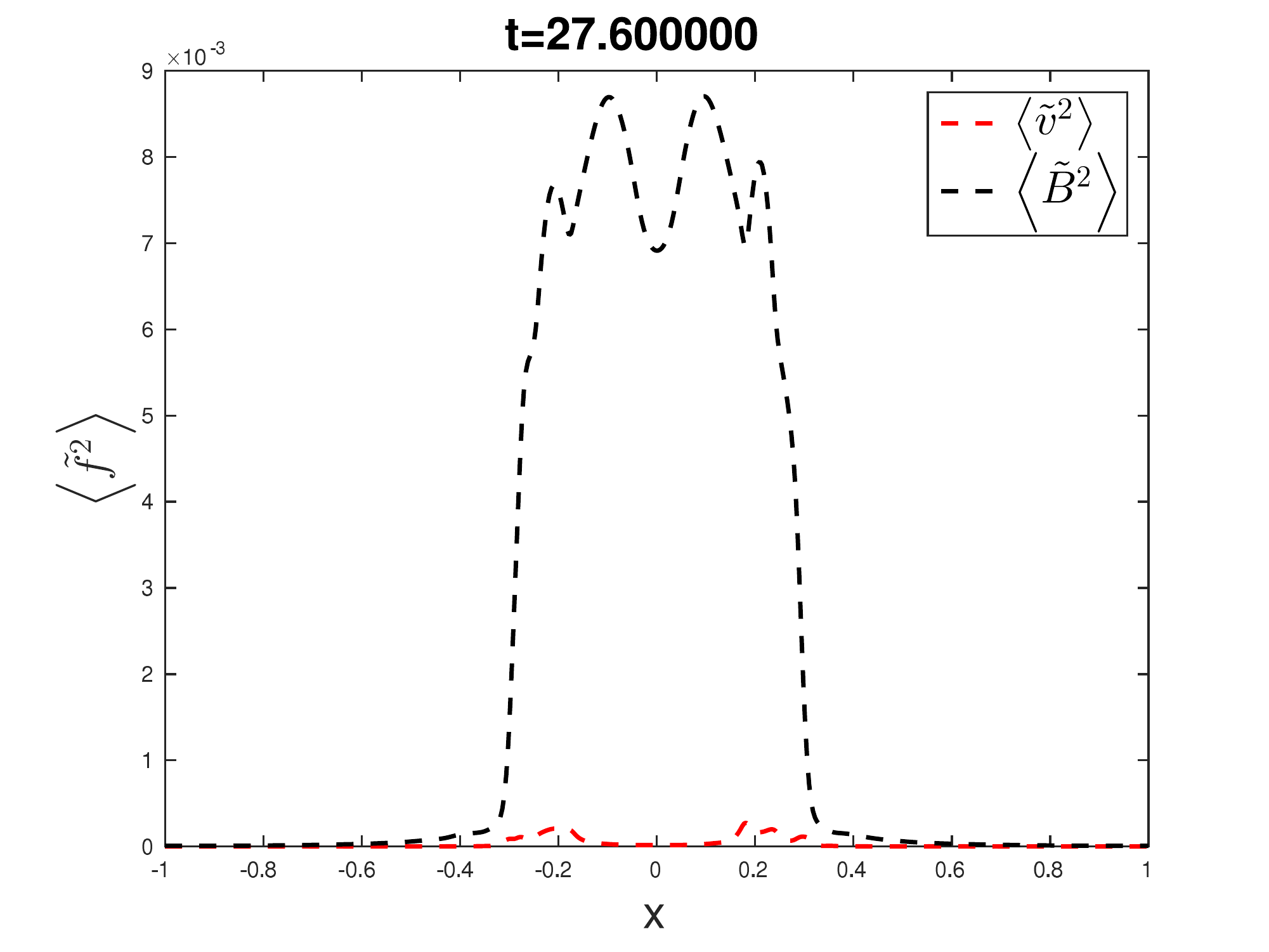}
}
\\
(a)&(b)
\\
\parbox{3.1in}{
  \includegraphics[scale=0.43]{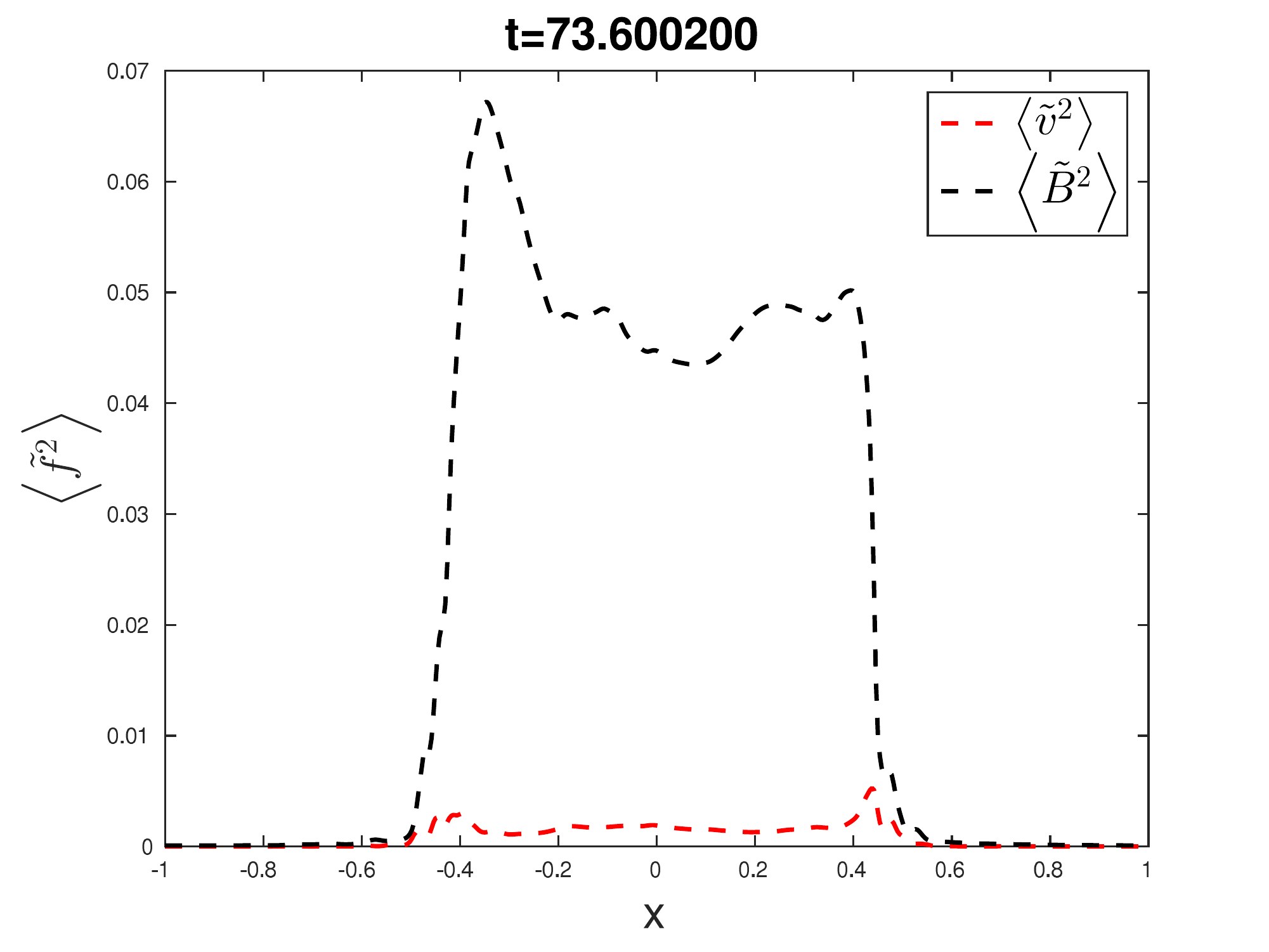}
}
&
\parbox{3.1in}{
  \includegraphics[scale=0.43]{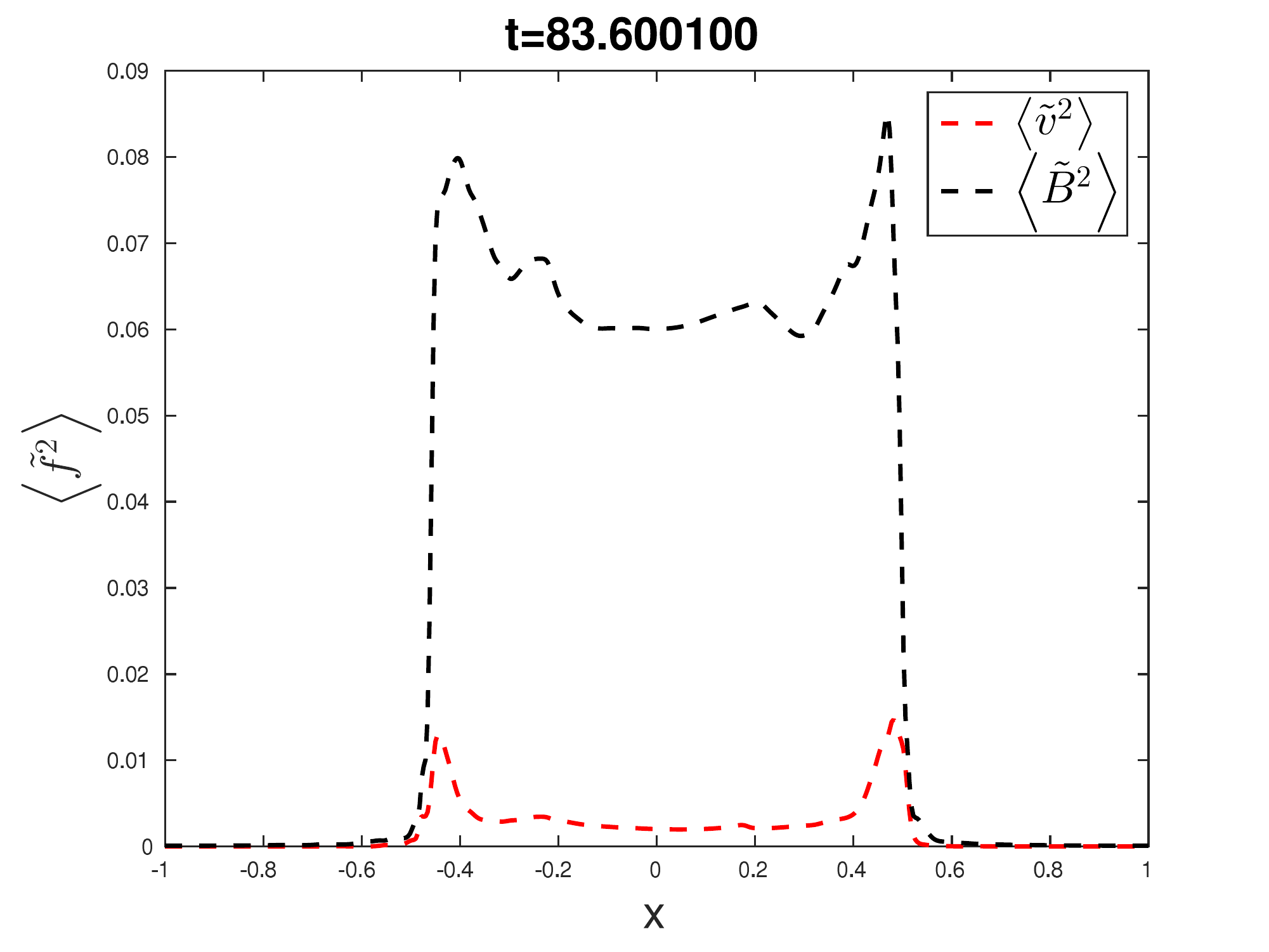}
}
\\
(c)&(d)
\etbl
\caption{The perturbed energy profile for different times. Most of the
  perturbed energy locates within $x\in\left[-0.5,0.5\right]$. At the
  beginning of the simulation, only a few large-scale unstable tearing modes
  exist, and the envelope of their mode structures
  defines the perturbed energy profile. As time goes on, these large-scale
  modes overlap with each other and generate the tearing
  turbulence, flattening the perturbed energy profile.
}
\label{fig:2}
\end{figure*}

In this subsection, we will compare the tearing turbulence in a strong
guide field with our previous theoretical model. Let $X=2$. $Y=4$,
$Z=20$, $B_{y0}\left(0\right)=1.5$ and $B_{0}=10$.

At the beginning of the simulation, small initial
perturbations with harmonics $m/n=3/2$, $2/1$ and $3/1$ are seeded. The resonant
surfaces corresponding to those modes lie in the central region of the
system $x\in\left[-0.5,0.5\right]$, as can be seen from
Fig.\,\ref{fig:1}. This is also the region where the turbulence
amplitude is strongest later in time. The
quadratic form of magnetic and kinetic perturbation is averaged across
the $y$-$z$ plane, providing us the sum of perturbation energy over
the whole spectrum, $\left<\tsv^2\right>$ and $\left<\taB^2\right>$, given by
\bqbl
\left<\taB^2\right>
=
\left<B^2\right>
-\left<B\right>^2
,\eqbl
\bqbl
\left<\tsv^2\right>
=
\left<v^2\right>
-\left<v\right>^2
.\eqbl
These perturbation energies as functions of $x$ are
plotted in Fig.\,\ref{fig:2} for different times. Initially, the dynamics
is dominated by a few large-scale unstable modes, and the envelope of
their mode structure determines the perturbation energy profile, as can
be seen from Fig.\,\ref{fig:2} (a) and Fig.\,\ref{fig:2} (b). Later,
the initial islands overlap with each other and generate a large spectrum of
small-scale modes, and the perturbation energy profile becomes
smooth in the core region, as seen in Fig.\,\ref{fig:2} (c) and
Fig.\,\ref{fig:2} (d). By the time the tearing turbulence enters quasi-steady state,
both the magnetic and kinetic energy perturbations are
confined within the region $x\in\left[-0.5,0.5\right]$, and their
profiles are almost flattened within the central region. The kinetic
perturbation seems to be much smaller than the magnetic
perturbation, which would appear to raise doubt regarding our energy equipartition assumption.
However, as will be seen later in Section \ref{s:StrongField},
this is because equipartition is established not at the scale of the
large-scale instabilities driving the turbulence but at the small scales.
Meanwhile, the spectrum
for high-$k$ modes actually agrees rather well with the equipartition
assumption.

\begin{figure*}
\centering
\noindent
\btbl{cc}
\parbox{3.2in}{
  \includegraphics[scale=0.43]{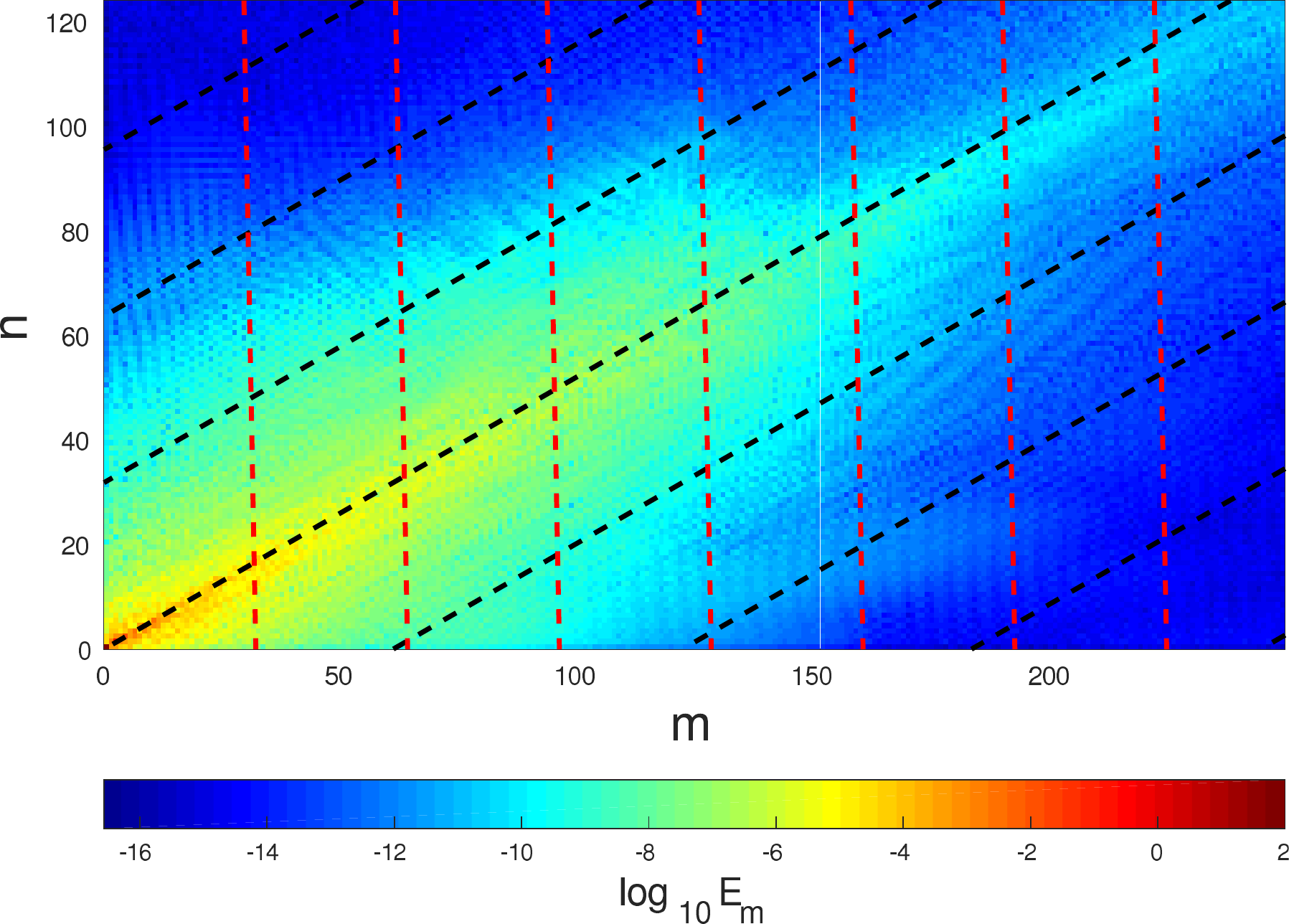}
}
&
\parbox{3.2in}{
  \includegraphics[scale=0.43]{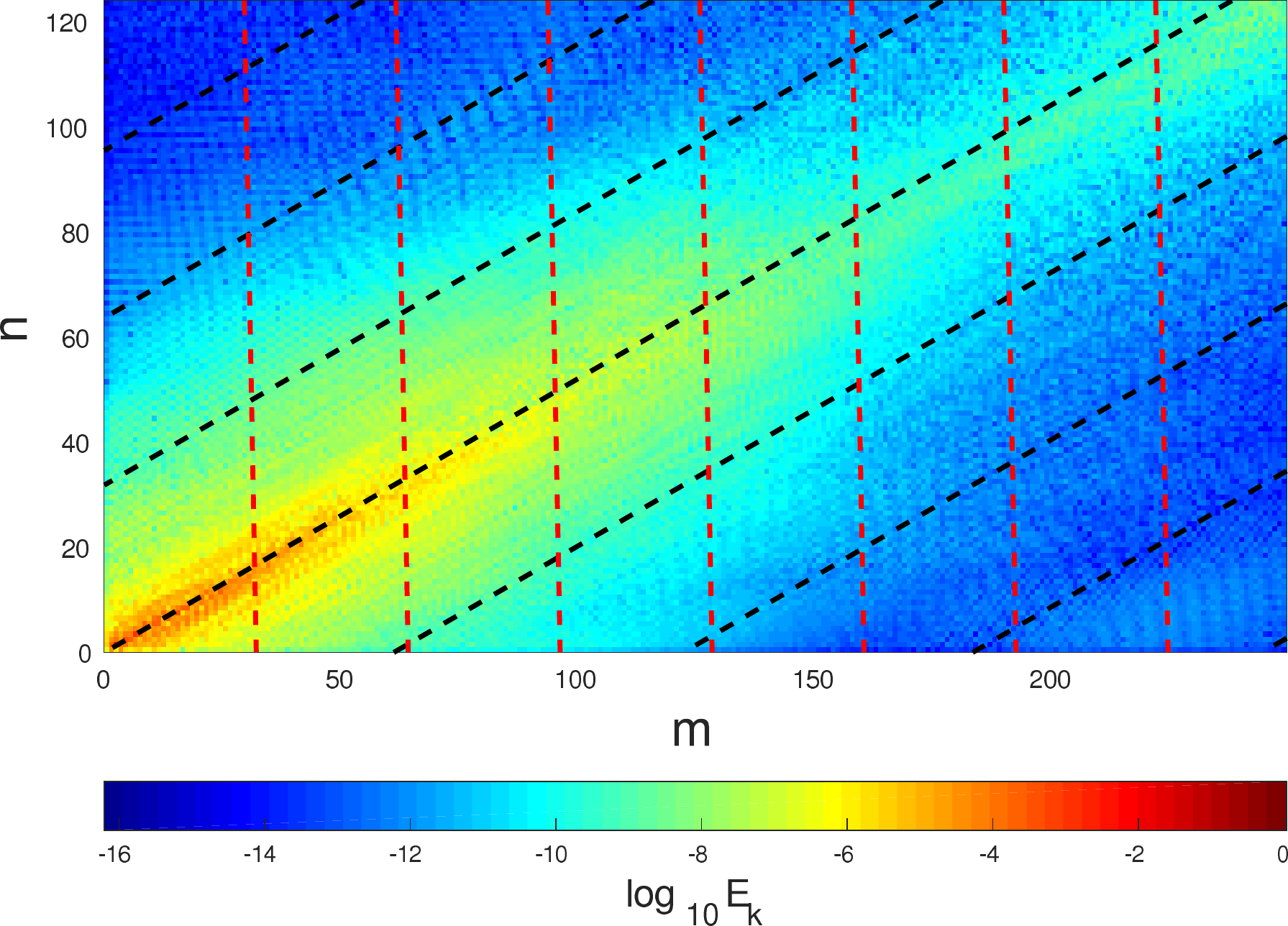}
}
\\
(a)&(b)
\etbl
\caption{The alignment of turbulent eddies to local mean field line at $x=0$. The
  color here represents the logarithm of perturbed energy density of (a)
  magnetic perturbation, and (b) kinetic perturbation. The black dashed
  lines are the contours of $\wsk\cdot\waB_0$. It can be seen that the
  turbulence is highly anisotropic, and tend to align with the direction
  of strong mean field. The red dashed lines
  represent the contours of $k_\bot$ in $\left(m,n\right)$ plane.
}
\label{fig:3}
\end{figure*}

The alignment of the turbulent eddies to the local mean field is also of
interest. That is, we wish to know whether or not the eddies have
elongated structure along the mean field direction, as would be expected
from highly magnetized MHD turbulence. Here, we look at the local
property of magnetic perturbation for a given $x$ position, and perform
Fourier decomposition along $y$ and $z$ direction for all components of
magnetic field. We take the zeroth order harmonic as the local mean
field for the given $x$ position, while all the other harmonics
correspond to modes with various scales. The alignment of those modes
to the direction of local mean field line can be represented by looking
at $\wsk\cdot\waB_0$. We once again write
\bqbl
\wsk\cdot\waB_0
=
B_{y0}k_y
-B_{z0}k_z
=
k_y\R{Y}{Z}\left(\gm-\R{n}{m}\right)B_{z0}
.\eqbl
Such alignment of small-scale tearing modes can then be checked by
looking at the distribution of the 2-D perturbed energy spectrum
$\left|\taB_k\right|^2$ and $\left|\tsv_k\right|^2$ in
$\left(m,n\right)$ space. The result for $x=0$ is shown in Fig.\,\ref{fig:3},
where the logarithm of the perturbed energy is plotted as a function of mode
number $m$ and $n$. The black dashed line represents the contour of
$\wsk\cdot\waB_0$, with the one originating from the $\left(0,0\right)$
point corresponding to $\wsk\cdot\waB_0=0$. It can be seen that the energy
spectrum strongly aligns with the local mean field, indicating a
strongly anisotropic structure.
It is noteworthy that, for a magnetically sheared system, this localization in the
$k$ space directly corresponds to the localization of mode structures near
their respective resonant surfaces in configuration space. Due to this
localized mode structure, the amplitude of the mode decreases rapidly as we move
away from its resonant surface. Thus, only modes which are near resonance
(corresponds to low $k_\|$) can be seen from the spectrum shown in
Fig.\,\ref{fig:3}, resulting in observed localization in the $k$ space. This is especially
true for high-$k$ modes. The red
dashed lines represent the contours of $k_\bot$. The strong alignment
behavior of small-scale perturbation in the presence of strong guide
field indicates that we have $k_\|\ll k_\bot$, and consequently
$k_y^2+k_z^2=k_\bot^2+k_\|^2\simeq k_\bot^2$. This confirms our previous
assumptions.

\begin{figure*}
\centering
\noindent
\btbl{c}
\parbox{5.5in}{
  \includegraphics[scale=0.48]{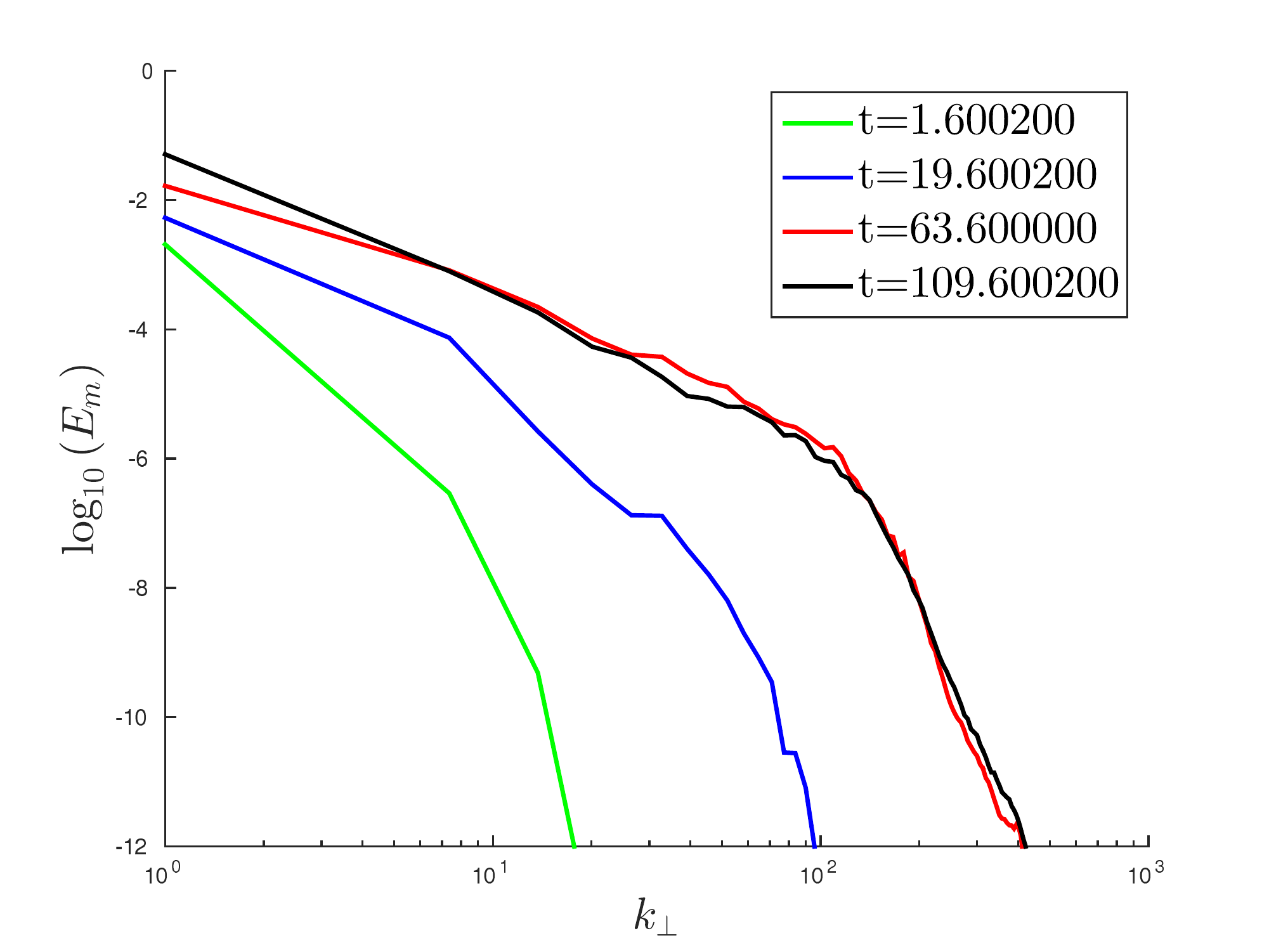}
}
\etbl
\caption{The time evolution of the tearing turbulence energy spectrum. At $t=0$, there are only several unstable modes.  These large scale modes gradually excite a spectrum of small-scale modes by interacting with each other. At a later time, the tearing turbulence reaches a quasi-steady state.
  }
\label{fig:4}
\end{figure*}

With the alignment of eddies known, we now look at how this highly
anisotropic turbulence establishes itself.
From Fig.\,\ref{fig:2} (c) and (d), it can be seen that the turbulence
strength is rather flat in the central region, this implies that we can
use the local spectrum for a given $x$ to represent the evolution of
global tearing turbulence. Here, we choose to look at the spectrum
evolution at $x=0$. The energy density in $k_\bot$ space
$E\left(k_\bot\right)$ can be obtained by integrating over the red dashed
lines in Fig.\,\ref{fig:3}. The spectrum of $E\left(k_\bot\right)$ for
several different times is presented in Fig.\,\ref{fig:4}. The logarithm
of magnetic energy perturbation is shown as a function of the logarithm
of the perpendicular scale $k_\bot$. It can be seen that at $t=0$ there are
only several large-scale unstable modes. Then the interaction of these
large-scale modes gradually stir up small-scale modes.
At a later time, the tearing turbulence reaches a quasi-steady state as can
be seen in Fig.\,\ref{fig:4}. The structure of $E\left(k_\bot\right)$ spectrum
changes very little from $t=63.6$ to $t=109.6$ while the longest
non-linear turnover time of the eddies is on the order of $\gt_{nl}\sim 1$.

\begin{figure*}
\centering
\noindent
\btbl{c}
\parbox{5.5in}{
  \includegraphics[scale=0.48]{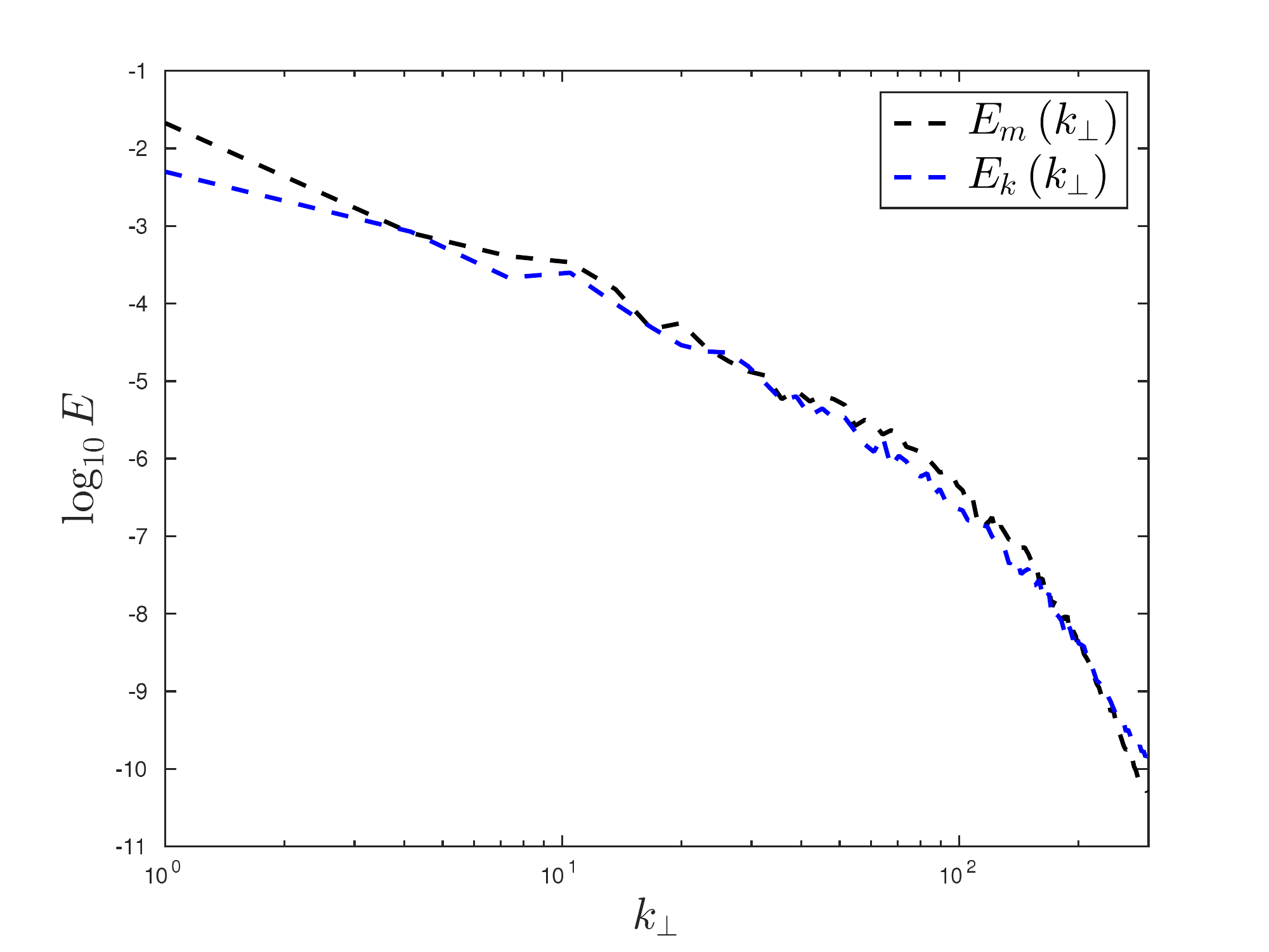}
}
\etbl
\caption{Comparison between kinetic and magnetic spectrum for
  $t=85.6$. It can be seen that the equipartition of energy is
  reasonably satisfied for high-$k_\bot$ modes, while there is some
  departure from equipartition for large-scale modes. The perturbed magnetic energy is approximately
  one order of magnitude larger than the kinetic energy for the largest
  mode.
}
\label{fig:5}
\end{figure*}

The comparison between magnetic and kinetic energy spectrum after the
turbulence reached the quasi-steady state is another important issue, as we have assumed an equipartition of energy in Section \ref{s:EnergyBudget}. An
example kinetic and magnetic spectrum for quasi-steady state tearing
turbulence is shown in Fig.\,\ref{fig:5} for $t=85.6$. It can be seen
that for high-$k$ modes the kinetic and magnetic energy are
approximately the same, while at the largest scale there is a departure
from equipartition. The departure does not significantly impact our
theoretical analysis in Section \ref{s:EnergyBudget}, since we are
primarily concerned with small-scale modes which are linearly stable
rather than the unstable large-scale modes. As a side note, the fact that the
magnetic perturbation is one order of magnitude larger than the
kinetic perturbation for largest scale modes is also consistent with the
observation in Fig.\,\ref{fig:2}, as the total magnetic perturbation
energy is also one order of magnitude larger than the total kinetic
perturbation energy.

The next important property we are interested in is the structure function of the turbulence,
which provides us information regarding the scale dependence of
its anisotropy and thus has significant impact on the energy transfer rate
and consequently the turbulence spectrum. We follow the procedure
detailed in Ref.\,[\onlinecite{Huang16APJ}] and
Ref.\,[\onlinecite{Cho2000}], and define the following two-point structure
functions:
\bqbl
F_k\left(l_\|,l_\bot\right)
\equiv
\left<\left|\wsv\left(\wgz+\wsl\right)-\wsv\left(\wgz\right)\right|^2\right>
,\eqbl
\bqbl
F_m\left(l_\|,l_\bot\right)
\equiv
\left<\left|\waB\left(\wgz+\wsl\right)-\waB\left(\wgz\right)\right|^2\right>
.\eqbl
Here, $\wgz=\left(x,y,z\right)$ is the position of a random point in the
configuration space, and $\wsl$ is a random vector. Thus,
$\wgz+\wsl$ and $\wgz$ define a random pair of points in configuration
space. The bracket $\left<f\right>$ here indicates an ensemble
average over a large number of random pairs. Due to the strong
localization of mode structure demonstrated in Fig.\,\ref{fig:3},
we look at a 2D version of the structure function in our study. That is,
we take the random pairs within the $y$-$z$ plane for a given $x$ instead
of considering the full 3D space. The parallel and perpendicular
component of $\wsl$ is defined by the local mean field direction, which
is calculated by averaging the magnetic field at two points. We
average over $10^9$ random pairs of points, and obtain the structure function
for both kinetic and magnetic perturbation as functions of $l_\|$ and
$l_\bot$. The contours of this structure function in
$\left(l_\|,l_\bot\right)$ then reflect the anisotropy of the eddy at
different $l_\bot$ scales.

\begin{figure*}
\centering
\noindent
\btbl{c}
\parbox{5.5in}{
  \includegraphics[scale=0.48]{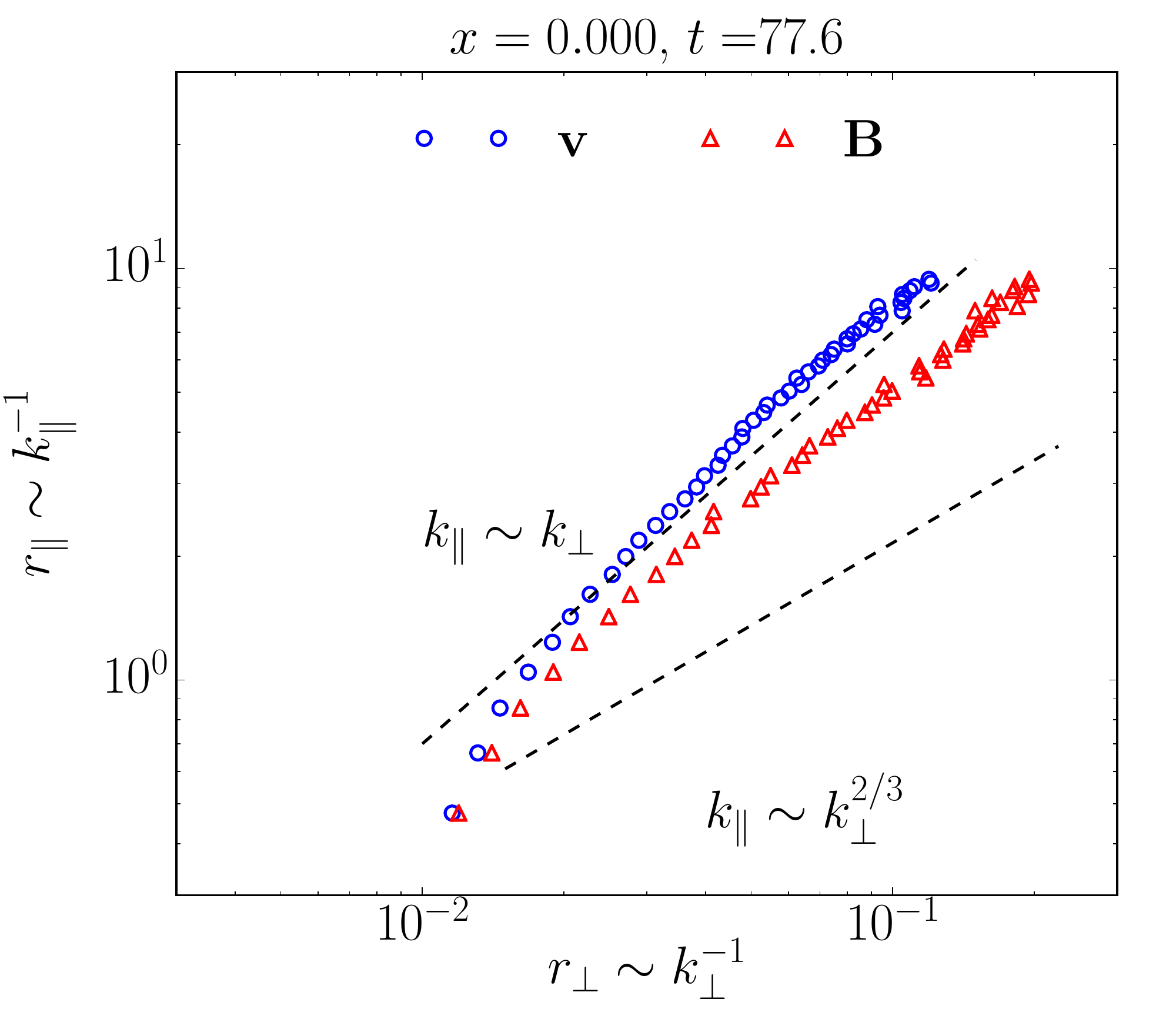}
}
\etbl
\caption{The anisotropy of tearing turbulence for various scale
  $l_\bot$. Both the kinetic and magnetic perturbation is shown. The
  scaling of both $k_\|\propto k_\bot$ and $k_\|\propto k_\bot^{2/3}$ are
  shown as black dashed lines. It can be seen that the turbulence
  anisotropy is scale independent $k_\|\propto k_\bot$ for most of
  the scales.
}
\label{fig:6}
\end{figure*}

To extract this anisotropy information, we search for the intersection of a
given contour of $F_k\left(l_\|,l_\bot\right)$ or $F_m\left(l_\|,l_\bot\right)$ with
the $l_\|$ and $l_\bot$ axis respectively. Thus we can obtain a pair of $l_\|$ and $l_\bot$
for a given contour of $F_k\left(l_\|,l_\bot\right)$ or $F_m\left(l_\|,l_\bot\right)$,
the ratio of which represents the anisotropy at a given scale. A scan of these $l_\|$ and $l_\bot$
pairs then shows the scale dependence of turbulence anisotropy.
The anisotropy thus obtained is plotted in
Fig.\,\ref{fig:6}, with two scalings $k_\|\propto k_\bot$ and $k_\|\propto
k_\bot^{2/3}$ plotted as black dashed lines.
It can be seen that the anisotropic behavior of simulation result
largely agrees with our analytical model and is mostly scale-independent. There is some discrepancy between the length scale of
kinetic and magnetic perturbations, which might be the consequence of
their different distribution width in $\left(k_y,k_z\right)$ space as can be
seen in Fig.\,\ref{fig:3}. The ratio
between parallel and perpendicular length scale ultimately deviates from
the scale-independent scaling at the very small scale where classical
dissipation kicks in. Lastly, it is observed that $l_\|$ is two orders of
magnitude larger than $l_\bot$, thus we hereby take $\ga\sim 10^{-2}$ as a
reasonable estimation. This estimation also agrees with our prediction by
Eq.\,(\rfq{eq:scaleindep}) since we also have
$\taB_L/B_{z0}\sim\mathcal{O}\left(10^{-2}\right)$.

\begin{figure*}
\centering
\noindent
\btbl{c}
\parbox{5.5in}{
  \includegraphics[scale=0.05]{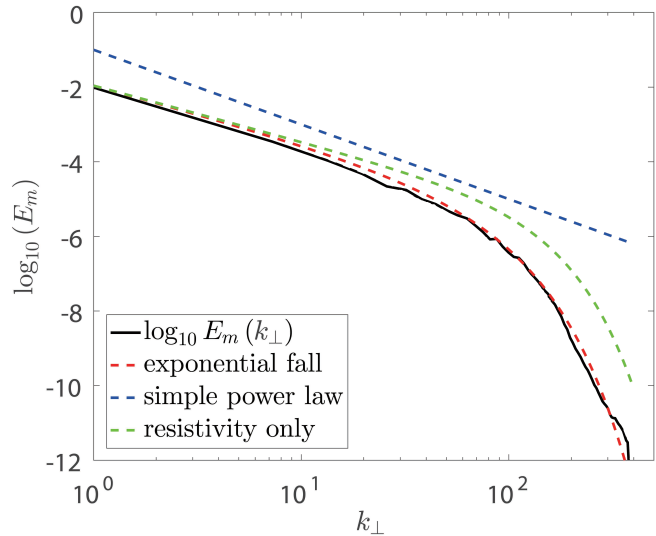}
}
\etbl
\caption{The spectrum of tearing turbulence for $t=85.6$ and $x=0$,
  compared with three analytical models. The black solid line represents the
  simulation result, the red dashed line is a fitting of
  Eq.\,(\rfq{eq:Spectrum}) using only the energy injection rate $\ge$ as
  a free parameter, the blue dashed line corresponds to a simple
  power law $E_m\left(k_\bot\right)\propto k^{-2.0}$, and the green dashed
  line is the spectrum produced by Eq.\,(\rfq{eq:Spectrum}) if we only consider
  the resistive damping. The simulation result agrees rather well with our
  analytical prediction.
}
\label{fig:7}
\end{figure*}

With the characteristic structure known, we can finally check our
analytical model given by Eq.\,(\rfq{eq:Spectrum}) against the simulation
results. The magnetic perturbation spectrum for tearing turbulence is
shown in Fig.\,\ref{fig:7} for $t=85.6$ and $x=0$.
The simulation result
is compared with three analytical models: our damped turbulence model as
shown in Eq.\,(\rfq{eq:Spectrum}), a simple power law
$E\left(k_\bot\right)\propto k_\bot^{\gb}$ as a result of the traditional
inertial range argument, and the spectrum produced by Eq.\,(\rfq{eq:Spectrum})
if only the resistivity is included as damping.
From numerical observation, we estimate the
characteristic length scale to be $\gl\simeq 10$ near the central flattened
region where the resonant surfaces of the concerned modes lie.
Here $\gl$ can be larger than $X$ since it only serves as an indication of
the local magnetic field gradient.
The only free parameter in Eq.\,(\rfq{eq:Spectrum}) is
then the energy injection rate $\ge$, which will be used to fit the
simulation result. On the other hand, the power index $\gb$ in the simple
power law will also be used as a free parameter to fit the numerical
result. The fitting exercise yields $\ge\simeq 1.2\times 10^{-3}$ and $\gb\simeq
-2.0$, with fixed parameters $\gl=10$ and $\ga=0.01$. The fitted curves
are shown in Fig.\,\ref{fig:7}. It can be seen that
our analytical model is in better agreement with the simulation result
than either the simple power law or the spectrum obtained by assuming that it is determined by the effect of resistivity only.
It is noteworthy that although the final decay of the turbulence spectrum
is due to the influence of resistivity, the actual curve deviates from the inertial range curve due to the presence of effective damping. While this deviation might suggest that there exists an inertial range with a steeper slope represented, for instance, by the blue dashed line, this is not the case since the behavior seen is caused by slow exponential decay and cannot be represented accurately by a power law.

\subsection{Weaker guide field case}
\label{s:WeakField}

\vskip1em

The strong guide field case has been investigated in the previous
subsection. Reasonable agreement has been found between the simulation
result and our theoretical prediction obtained in Section
\ref{s:DampedTurbulence}. The magnitude of the perturbation has been found to be two order of
magnitude smaller than the guide field. However, we are also interested in cases
where the guide field is weaker, and the large-scale Kubo number
$\gk=\left(\taB_L/B_{0}\right) \left(L_\|/L_\bot\right)\simeq
1$. Again, $\gk$ here is equivalent to the $\gc$ used in
Ref.\,[\onlinecite{GS1997}]. Note that, in this case of stronger turbulence, the
perturbed field is still smaller than the guide field, although the Kubo
number may exceed unity due to anisotropy.

The initial magnetic fields are now $B_{y0}\left(0\right)=1.5$ and
$B_{0}=2.5$. To maintain a similar initial safety factor profile with
the one shown in Fig.\,\ref{fig:1}, the system size is now $X=2$, $Y=4$
and $Z=4$. We are mainly concerned with the anisotropic behavior and the
energy spectrum of the turbulence, and we wish to determine whether or not
the distinctive features exhibited in our weak turbulence simulation
persist in this stronger turbulence case.

\begin{figure*}
\centering
\noindent
\btbl{c}
\parbox{5.5in}{
  \includegraphics[scale=0.48]{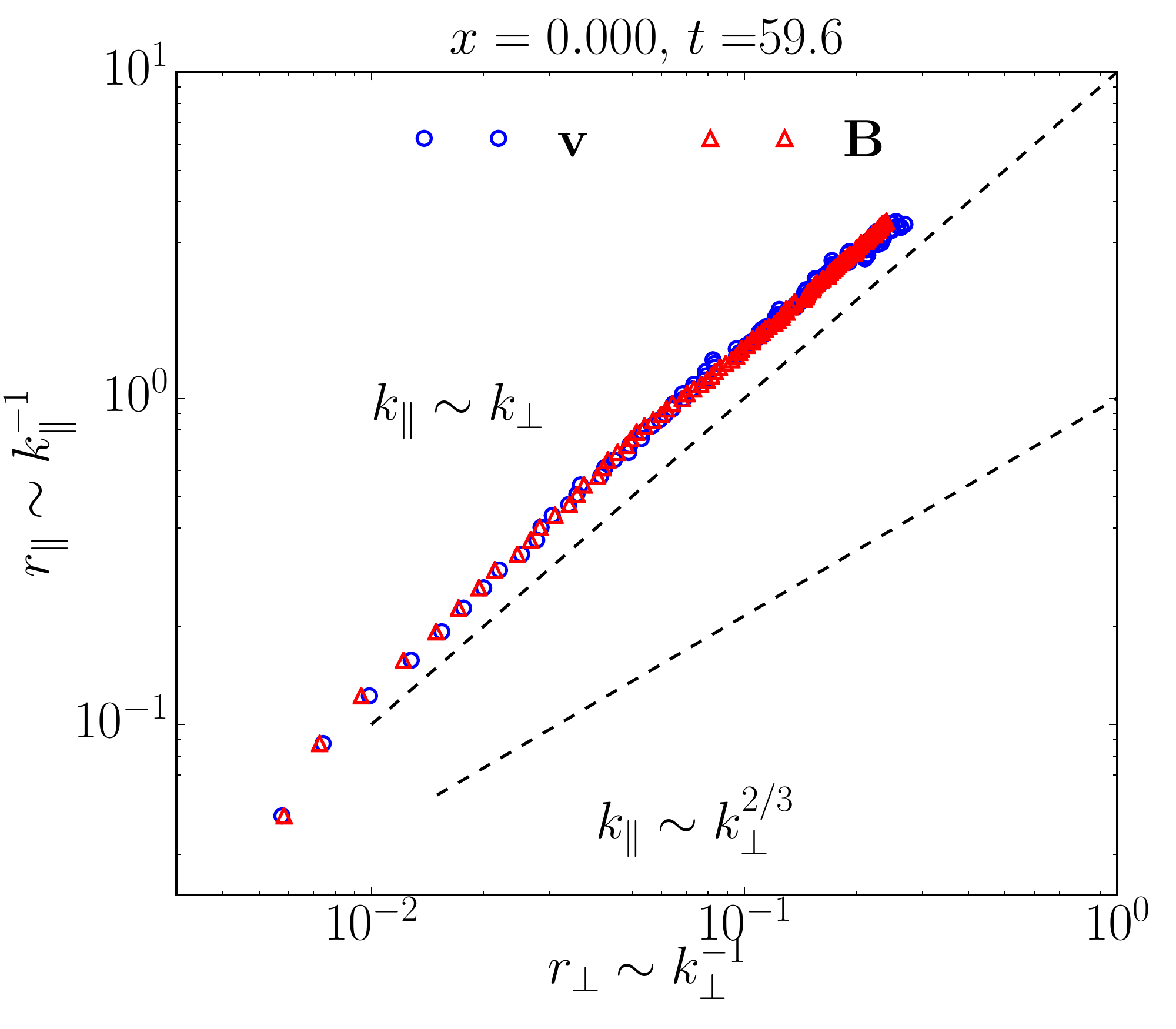}
}
\etbl
\caption{The scale-independent anisotropy for a weaker guide field
  case. Although the turbulence is now less anisotropic than in the
  strong guide field case, the fundamental scale-independent
  anisotropy remains the same comparing with what is shown in
  Fig.\,\ref{fig:6}.
}
\label{fig:8}
\end{figure*}

We first examine the anisotropy. Again, we look at the contours
of structure functions for both the kinetic and magnetic perturbation as
described in Section \ref{s:StrongField}, and we use the same technique
detailed there to extract the turbulence anisotropy for different scales. The
$x$ position is chosen at $x=0$, and time $t=59.6$, when the
turbulence has already reached the quasi-steady state. The
anisotropy is shown in Fig.\ref{fig:8}, with the two scalings $k_\|\propto
k_\bot$ and $k_\|\propto k_\bot^{2/3}$ plotted as black dashed lines. It
can be seen that this stronger turbulence case still follows the scale-independent anisotropy behavior described in Section \ref{s:Anisotropy} and only deviates from it at very small scales. In fact, the scale-independent anisotropy is even better compared to that shown in Fig.\,\ref{fig:6}, possibly due to a stronger large-scale perturbation.

\begin{figure*}
\centering
\noindent
\btbl{c}
\parbox{5.5in}{
  \includegraphics[scale=0.05]{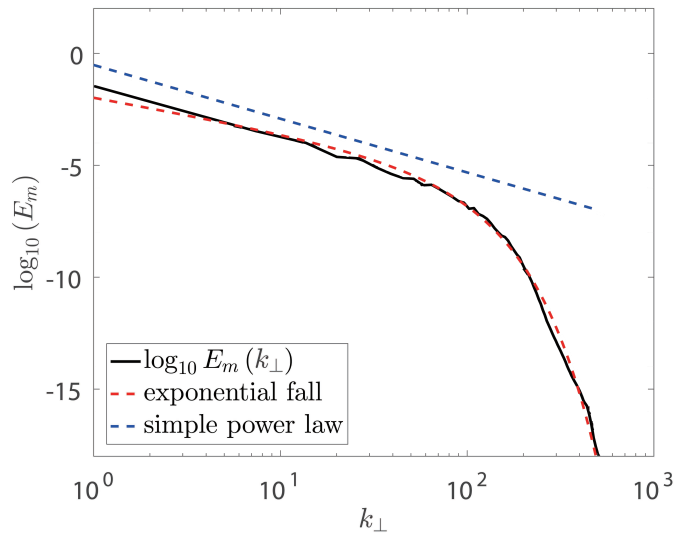}
}
\etbl
\caption{The spectrum of tearing turbulence for $t=59.6$ and $x=0$,
  compared with two analytical model. As is shown in Fig.\,\ref{fig:7},
  the simulation result agrees rather well with our analytical
  prediction, suggesting that our analytical model works well even for
  not-so-weak turbulence.
}
\label{fig:9}
\end{figure*}

We then look at the structure of the turbulence spectrum. The magnetic
perturbation spectrum for $x=0$ and $t=59.6$ is shown in
Fig.\ref{fig:9}. Once again, the simulation result is compared with a
simple power law $E\left(k_\bot\right)\propto k_\bot^{-2.4}$ and the
spectral form predicted by our model as described by Eq.\,(\rfq{eq:Spectrum}),
with fixed parameters $\gl=10$ and $\ga=0.05$. The fitting result
returns $\ge=8.5\times 10^{-4}$. Reasonable agreement is again found between
the numerical result and our prediction, with a gradual departure from the
original $k^{-3/2}$ scaling well before entering the resistive dissipation scale.

A noteworthy feature of this stronger turbulence case is that the Kubo
numbers for the largest perturbations exceed unity. From
Fig.\,\ref{fig:8}, it can be seen that the scale-independent anisotropy
is approximately $l_\|/l_\bot\simeq 20$. At the same time, numerical
observation from Fig.\,\ref{fig:9} indicates the largest scale
perturbation has
$\taB_L/B_{0}\sim\mathcal{O}\left(10^{-1}\right)$. Hence, for the large-scale perturbations, we have $\gk\simeq 2$, while for smaller scale
perturbation the Kubo number steadily decreases as the perturbation
strength decreases. This is different from the critical balance scenario
where the Kubo number remains on the order of unity across all scales.
This deviation from critical balance is very similar to
that discussed by Huang et al. in Ref.\,[\onlinecite{Huang16APJ}]. Thus, we
conclude that our analysis can also be applied to the case where the
nonlinear mixing is stronger than the linear parallel propagation, such
as those reported in plasmoid turbulence, where
the critical balance condition was frequently assumed to be true.

\section{Discussion and conclusion}
\label{s:Conclusion}

\vskip1em

Instability driven tearing turbulence in sheared magnetic field is
studied in this work. The turbulence consists of several large-scale
unstable modes and a broad spectrum of small-scale linearly stable modes
which are excited by their large-scale counterparts.
It is found that the linearly stable modes will act as an effective
damping mechanism which has a weaker dependence on $k_\bot$ than classical dissipation.
For inviscid and viscous regimes, the dependence scales as $k_\bot^{6/5}$ and $k_\bot^{4/3}$ respectively.
The weak dependence indicates that this damping mechanism will manifest itself long before turbulence eddies
reach the resistive or viscous dissipation scales. Consequently, a well-defined inertial range cannot be identified, and damping must be considered at all scales.
Furthermore, we argue that the tilting of sheared background field by large-scale
perturbations will impose a scale-independent anisotropy for small-scale modes. This anisotropic behavior then determines the scale dependence
of the forward energy cascade rate.

With the knowledge of effective damping rate and energy cascade rate at
hand, the structure of this damped turbulence can be obtained by
considering local energy budget in the $k_\bot$ space. The key idea is that the
difference of energy forward transfer rate between the two ends of any
interval in the $k_\bot$ space corresponds to the damping within that
interval. The resulting spectrum features a power law multiplied by an
exponential falloff, as opposed to the pure power-law spectrum
obtained by using the standard inertial range argument.

The above analytical result is checked against visco-resistive MHD
simulations. The turbulence is found to be highly anisotropic and tends
to align with the strong local mean-field direction. The two-point
structure functions are calculated to investigate anisotropic property at
different scales, and a scale-independent anisotropy is found,
confirming our $l_\|\propto l_\bot$ argument. Furthermore, the equipartition
between kinetic and magnetic energy is found to be valid for the
turbulence in question. The numerical result appears to
agree well with our analytical model based on effective damping.

The behavior of a stronger turbulence, where the Kubo number exceeds
unity for certain scales, is also investigated. We find that the scale-independent anisotropy and the energy spectrum continue to hold
for the stronger turbulence case, indicating that our analysis remains
applicable even for the scenario where the perpendicular
turbulence shearing is stronger than the parallel propagation.

With this knowledge regarding spectrum structure, the next step would be
considering the back-reaction of small-scale turbulence on large scales. This involves a sum of quadratic form of perturbed quantities over the whole
spectrum, which requires knowledge regarding the form of spectrum given
by our study here. An example is the small-scale spreading of mean field
described by hyper-resistivity, as has been studied in Ref.\,[\onlinecite{Strauss1986}]
and Ref.\,[\onlinecite{Craddock1991}]. Our analysis here provides a solid basis for future study along these lines. These studies are left to future work.

\vskip1em
\centerline{\bf Acknowledgments}
\vskip1em

  The authors thank P. H. Diamond, X.-G. Wang, H.-S. Xie and L. Shi for fruitful discussion. This work is partially supported by the National Natural Science Foundation of China under Grant No. 11261140326 and the China Scholarship Council. A. Bhattacharjee and Y.-M. Huang acknowledge support from NSF Grants AGS-1338944 and AGS-1460169, and DOE Grant DE-SC0016470. Simulations were performed with supercomputers at the National Energy Research Scientific Computing Center. D. Hu publishes this paper while working in ITER Organization. ITER is a Nuclear Facility INB-174. The views and opinions expressed herein do not necessarily reflect those of the ITER Organization.

\end{document}